\documentclass[twocolumn,  aps, physrev,amsmath,amssymb]{revtex4-2}
\usepackage{graphicx,bm}
\usepackage{color}
\usepackage{soul}
\usepackage{cases}
\usepackage{appendix}

\begin{document}

\preprint{APS/123-QED}

\title{Photon transport and blockade based on non-Markovian interactions between a microring resonator and waveguide}

\author{Haijin Ding} \email{dhj17@tsinghua.org.cn}

\affiliation{Department of Mechanical and Automation Engineering, The Chinese University of Hong Kong, Hong Kong}

\begin{abstract}
We investigate photon transport and blockade based on the architecture where a waveguide is coupled to a microring resonator at two distinct points. This two-point coupling configuration between the waveguide and resonator gives rise to non-Markovian dynamics, which is induced by the photon transmission delay in the waveguide between the two coupled points. On one hand, by designing the non-Markovian coupling parameters between the waveguide and resonator, single- or two-photon transport and the resulting photon blockade effect can be manipulated according to the output photonic states at the end of waveguide. We for the fist time evaluate the occurrence of photon blockade with scattering matrices and second order correlation functions based on this non-Markovian proposal related to the length of waveguide between two coupled points. On the other hand, when classical driving fields are applied upon the resonator with interactions between its clockwise and counterclockwise modes, the blockade effect of the output field can be determined by the intracavity eigenstates. Then the correlations of the output field as well as the intracavity states can be modulated by the non-Markovian coupling between the microring resonator and waveguide.
\end{abstract}
\maketitle

\section{Introduction} \label{sec:introduction}
Quantum photonic integrated circuits (QPICs) have emerged as a promising platform for realizing scalable quantum information processing (QIP) by manipulating photons with linear optical devices such as beam splitters and phase shifters~\cite{mower2015high,elshaari2020hybrid,knill2001scheme,wang2020integrated}.  The advantage of QPICs lies in the weak coupling of photons to the environment compared with other quantum systems such as atoms or ions, making it feasible to realize room-temperature QIP with photonic circuits~\cite{knill2001scheme}. Additionally, by incorporating nonlinear devices for the control of multi-photon states, QPICs can enable the realization of photonic Fock states and the generation of entanglements among photons~\cite{caspani2017integrated,jin2014chip}. Recent advancements have shown that on-chip photonic circuits can be utilized to create quantum states~\cite{ciampini2016path}, perform gate operations~\cite{politi2008silica,carolan2015universal} and algorithms for quantum computation~\cite{politi2009shor,tabia2016recursive}, realize quantum key distribution~\cite{ding2017high,sibson2017chip,paraiso2021photonic}, simulate molecule dynamics~\cite{zhang2019electronically} and so on. Thus QPICs have significant  potential applications in large scale quantum information processing due to their advantages in scalability and integrability~\cite{wang2020integrated}.

The generation of single-photon states is a crucial aspect in the realization of quantum information processing based on photonic integrated circuits~\cite{TwoSMwaveguideTwoCav}. Single-photon states can be generated through spontaneous emission when a quantum emitter is in free space~\cite{kimble1977photon}, coupled to a waveguide~\cite{zhao2020single}, within a cavity~\cite{mucke2013generation}, or modulated by nonlinear materials~\cite{peyronel2012quantum}. The photon blockade effect, which can suppress multiphoton events by blocking the generation of a second photon, has been widely applied to enhance the reliability of on-demand single-photon sources~\cite{liu2014blockade,wang2015tunable,tang2021towards,shen2024unconventionalYi}. For instance, in the experimental realization of low-temperature atom-cavity systems, the significant nonlinear interactions between the atom and a Fabry-Perot cavity can induce the blockade effect, thereby influencing the transmission of photons in the cavity~\cite{birnbaum2005photon}. Consequently, there can be more or less than one photon detected at the output side of the cavity. This experimental setup can be similarly generalized to the platforms based on microring resonators, which are commonly employed in the room-temperature quantum information processing~\cite{katsumi2025high,steiner2021ultrabright}. Similar to a Fabry-Perot cavity, a microring resonator can be coupled to atoms~\cite{aoki2006observation,dayan2008photon,liu2022proposal,yan2023unidirectional}, waveguides used for the mediation of photons~\cite{ShiTaoPRA,hach2010fully},  and Kerr materials with nonlinearity~\cite{SmatrixCal,TwoPhotonKerr,yuan2024phase}, resulting in linear and nonlinear quantum dynamics~\cite{miranowicz2013two,scott2019scalable,li2020photon}.

Moreover, non-Markovian quantum photonics can arise in systems where there are multiple microring resonators coupled to a waveguide or single microring resonator coupled to a waveguide at different points, due to the transmission delays of photons in the waveguide~\cite{banic2024integrated}. For example, when one resonator is coupled to a waveguide at two distinct points, the photons transmitted between two coupled points can construct a coherent feedback channel~\cite{OEinputOutput,zhou2007electrically,wang2016dynamicsSR}. In such non-Markovian networks, the two-point coupling between the waveguide and resonator is influenced by different photonic modes in the waveguide. For each single mode, the phase-modulated interactions between the resonator and waveguide is similar to the Markovian interactions between a resonator and environment~\cite{TwoSMwaveguideTwoCav}. In Markovian quantum systems,  especially for the circumstance with more than one photon,  a most precise approach to analyze the photon transportation and blockade is calculating the scattering matrices~\cite{fan2010input,SmatrixCal,xu2015input}. However, this approach has not be generalized to non-Markovian quantum networks with more than one coupling points between the resonator and waveguide.

In this paper, we for the first time introduce the approach in Refs.~\cite{fan2010input,SmatrixCal,xu2015input} to a quantum system with non-Markovian properties, realized by a waveguide coupled to a microring resonator at two distinct points. The photon transport and blockade can be manipulated by coupling strengths between the waveguide and resonator, the transmission delay of photons in the waveguide between the two coupled points, and the nonlinear coupling between the resonator and Kerr materials. The remainder of this paper is organized as follows. In Sec.~\ref{sec:model}, we derive the non-Markovian interaction model between the microring resonator and waveguide as a non-Markovian input-output formalism. Based on this, in Sec.~\ref{Sec:Scattering}, we study the non-Markovian transport of single and two photons evaluated by the scattering matrix, as well as the possibility of the occurrence of photon blockade. In Sec.~\ref{Sec:PBmasterEquation}, we study the non-Markovian dynamics when the resonator is driven by classical control fields, and the resulting blockade effect related with the two-, single- and zero-photon states. Finally, conclusions are presented in Sec.~\ref{Sec:conclusion}.

\section{Non-Markovian interactions between microring resonator and waveguide}\label{sec:model}
\begin{figure}[h]
\centerline{\includegraphics[width=0.9\columnwidth]{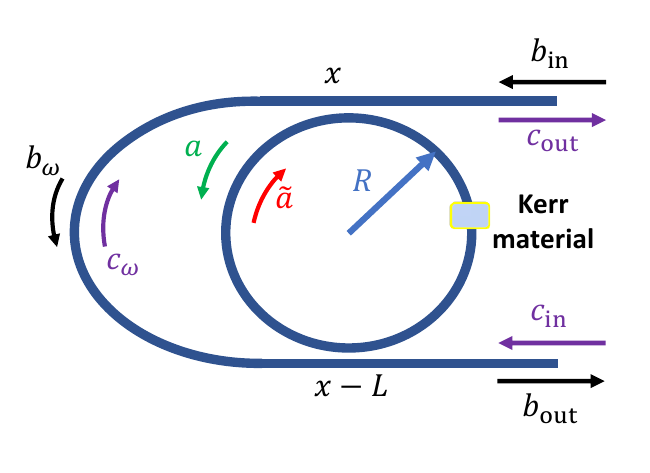}}
\caption{Microring resonator coupled to a waveguide at two distinct points.}
	\label{fig:system}
\end{figure}
As shown in Fig.~\ref{fig:system}, a microring resonator with the radius $R$ is coupled to a waveguide at two points $x$ and $x-L$, and the resonator is also coupled to a Kerr material. The overall Hamiltonian of the system can be represented by denoting $\hbar = 1$ as~\cite{fan2010input}
\begin{equation} \label{con:Htotal}
\begin{aligned}
&H =  \left(\omega_a - i\kappa_a \right) \left(a^{\dag}a+\tilde{a}^{\dag}\tilde{a}\right)+ H_{\rm w} + H_{\rm Kerr} + H_{\rm I},
\end{aligned}
\end{equation}
where the first component at the righthand side corresponds the Hamiltonian of a microring resonator with the resonant frequency $\omega_a$ and loss rate $\kappa_a$, $\tilde{a}$ ($\tilde{a}^{\dag}$) are the annihilation(creation) operators of the clockwise resonator mode, and $a$($a^{\dag}$) are those for the counter-clockwise resonator mode. The second component in Eq.~(\ref{con:Htotal}) represents the waveguide Hamiltonian as
\begin{equation} \label{con:Hw}
\begin{aligned}
H_{\rm w} = \int_{0}^{\infty}  \omega b_{\omega}^{\dag}b_{\omega}\mathrm{d}\omega + \int_{0}^{\infty}  \omega c_{\omega}^{\dag}c_{\omega} \mathrm{d}\omega,
\end{aligned}
\end{equation}
where $b_{\omega}$($b^{\dag}_{\omega}$) are the annihilation(creation) operators for the waveguide modes propagating from $x$ to $x-L$, $c_{\omega}$($c^{\dag}_{\omega}$) are those for the waveguide modes propagating from $x-L$ to $x$, and $\left [ b_{\omega}\left(\omega\right),b_{\omega}^{\dag}\left(\tilde{\omega}\right)\right] = \left [ c_{\omega}\left(\omega\right),c_{\omega}^{\dag}\left(\tilde{\omega}\right)\right] = \delta\left(\omega - \tilde{\omega}\right)$. The third component in Eq.~(\ref{con:Htotal}) is for the interaction between the microring resonator and the Kerr material with the strength $\chi$, and can be represented as~\cite{TwoSMwaveguideTwoCav,Jinghui2024chiral,yuan2024phase,zhou2022decay,zhong2021control,ChangchunPRA2024} 
\begin{equation} \label{con:Hkerr}
\begin{aligned}
H_{\rm Kerr} = \chi\left( a^{\dag}a^{\dag}aa +  \tilde{a}^{\dag}\tilde{a}^{\dag}\tilde{a}\tilde{a}+ 2a^{\dag}a\tilde{a}^{\dag}\tilde{a}\right).
\end{aligned}
\end{equation}
The last component of Eq.~(\ref{con:Htotal}) represents the remaining interaction Hamiltonian, including the interaction between the clockwise and counter-clockwise resonator modes, and the interactions between the resonator and waveguide, namely~\cite{zhang2020enhancement}
\begin{equation} \label{con:Hintj2nonchiral}
\begin{aligned}
&H_{\rm I} =\zeta a^{\dag} \tilde{a} + \zeta^*a \tilde{a}^{\dag} \\
&+  \int_{0}^{\infty} \left [ i e^{\frac{i\omega x}{c}} \left(\sqrt{\frac{g_1}{2\pi}}   + \sqrt{\frac{g_2}{2\pi}} e^{-\frac{i\omega L}{c}} \right) b^{\dag}_{\omega} a +{\rm H.c.} \right]  \mathrm{d}\omega\\
&+  \int_{0}^{\infty}\left [ i e^{-\frac{i\omega x}{c}} \left(\sqrt{\frac{\tilde{g}_1}{2\pi}}  + \sqrt{\frac{\tilde{g}_2}{2\pi}} e^{\frac{i\omega L}{c}} \right) c^{\dag}_{\omega} \tilde{a} +{\rm H.c.} \right] \mathrm{d}\omega,
\end{aligned}
\end{equation}
where $\zeta$ is the intermodal coupling strength between two modes $a$ and $\tilde{a}$ in the resonator induced by surface roughness~\cite{ShiTaoPRA}, $g_1$ and $g_2$ represent the real-value coupling strengths between the resonator and the waveguide mode $b_{\omega}$ at $x$ and $x-L$ respectively, $\tilde{g}_1$ and $\tilde{g}_2$ represent those for the waveguide mode $c_{\omega}$ at $x$ and $x-L$ respectively, and $\rm{H.c.}$ is for Hermitian conjugate. 

Based on the Hamiltonian in Eq.~(\ref{con:Htotal}), the dynamics for an arbitrary operator $O$ is governed by $\dot{O} = -i [O,H]$. Then the dynamics of the system operators can be derived as follows,
\begin{small}
\begin{subequations} \label{eq:operatorDMaadd}
\begin{numcases}{}
\dot{a} = -i \omega_a a - \kappa_a a - i\zeta\tilde{a} - 2i\chi a^{\dag}a^2  -2i\chi a \tilde{a}^{\dag}\tilde{a} \notag \\
~~~~~- \int_{0}^{\infty}  \left (\sqrt{\frac{g_1}{2\pi}} e^{-  \frac{i\omega x}{c}} + \sqrt{\frac{g_2}{2\pi}} e^{-i\omega  \frac{x-L}{c}}\right )b_{\omega}\mathrm{d}\omega,\label{con:aevolution}\\
\dot{\tilde{a}}  = -i \omega_a \tilde{a} - \kappa_a \tilde{a} - i\zeta^*a  - 2i\chi \tilde{a}^{\dag}\tilde{a}^2 -2i\chi a^{\dag}a \tilde{a}\notag \\
~~~~~ - \int_{0}^{\infty}  \left (\sqrt{\frac{\tilde{g}_1}{2\pi}} e^{i\omega  \frac{x}{c}} + \sqrt{\frac{\tilde{g}_2}{2\pi}} e^{i\omega  \frac{x-L}{c}}\right )c_{\omega}\mathrm{d}\omega, \label{con:atidleevolution}\\
\dot{b}_{\omega}= -i\omega b_{\omega} +\left (\sqrt{\frac{g_1}{2\pi}} e^{i\omega \frac{x}{c}} + \sqrt{\frac{g_2}{2\pi}} e^{i\omega \frac{x-L}{c}}\right )a,\label{con:dkevolution}\\
\dot{c}_{\omega} =-i\omega c_{\omega} +\left (\sqrt{\frac{\tilde{g}_1}{2\pi}} e^{-i\omega  \frac{x}{c}} + \sqrt{\frac{\tilde{g}_2}{2\pi}} e^{-i\omega  \frac{x-L}{c} }\right )\tilde{a}.\label{con:ckevolution}
\end{numcases}
\end{subequations}
\end{small}%

For the waveguide mode $b_{\omega}$, according to the input-output formalism, the input field can be defined according to the initial state of the waveguide as~\cite{fan2010input}
\begin{subequations} \label{eq:FreqOperator}
\begin{numcases}{}
b_{\rm in}(t) = \frac{1}{\sqrt{2\pi}} \int_{-\infty}^{\infty}   b_{\omega}(\omega,t_0) e^{-i\omega \left(t-t_0\right)}\mathrm{d}\omega, \label{bindef}\\
b_{\omega}(\omega,t_0) = \frac{1}{\sqrt{2\pi}} \int_{-\infty}^{\infty}  b_{\rm in}(t) e^{i\omega \left(t-t_0\right)}\mathrm{d}t, \label{dwt0def}
\end{numcases}
\end{subequations}
where $t_0$ represents the initial time for evolutions. Then $b_{\omega}(\omega,t)$ can be solved as
\begin{equation} \label{con:domegat}
\begin{aligned}
b_{\omega}(\omega,t) = \frac{1}{\sqrt{2\pi}} \int_{t_0}^{t} b_{\rm in}(t') e^{i\omega \left(t'-t\right)}\mathrm{d}t'.
\end{aligned}
\end{equation}
Thus Eq.~(\ref{con:aevolution}) can be equivalently represented with the input field $ b_{\rm in}(t)$ as
\begin{equation} \label{con:adynamics}
\begin{aligned}
\dot{a}=& -i \omega_a a  - \kappa_a a  - i\zeta\tilde{a} - 2i\chi a^{\dag}a^2 -2i\chi a \tilde{a}^{\dag}\tilde{a} \\
&- \left [\sqrt{g_1} b_{\rm in} \left(t+\frac{x}{c}\right) + \sqrt{g_2} b_{\rm in}\left(t+  \frac{x-L}{c}\right)\right ],
\end{aligned}
\end{equation}
where the cavity operator dynamics is influenced by its coupling to the input filed in the waveguide with the strengths $g_1$ and $g_2$. The last delayed component in Eq.~(\ref{con:adynamics}) is due to the two-point coupling between the resonator and waveguide, thus $a(t)$ is not only influenced by the input field $b_{\rm in}(t)$, but also influenced by the historic input field $b_{\rm in}(t-L/c)$, which makes it different from the traditional Markovian circumstance such as in Refs.~\cite{TwoSMwaveguideTwoCav,SmatrixCal}.

Similarly, another input field $c_{\rm in}(t)$ can be defined according to the waveguide mode $c_{\omega}$ as
\begin{subequations} \label{eq:FreqOperatorC}
\begin{numcases}{}
c_{\rm in}(t) = \frac{1}{\sqrt{2\pi}} \int_{-\infty}^{\infty}  c_{\omega}(\omega,t_0) e^{-i\omega \left(t-t_0\right)}\mathrm{d}\omega,\\
c_{\omega}(\omega,t_0) = \frac{1}{\sqrt{2\pi}} \int_{-\infty}^{\infty} c_{\rm in}(t) e^{i\omega \left(t-t_0\right)}\mathrm{d}t,
\end{numcases}
\end{subequations}
with
\begin{equation} \label{con:comegat}
\begin{aligned}
c_{\omega}(\omega,t) = \frac{1}{\sqrt{2\pi}} \int_{t_0}^t c_{\rm in}(t') e^{i\omega \left(t'-t\right)}\mathrm{d}t'.
\end{aligned}
\end{equation}
Then the resonator mode $\tilde{a}(t)$ in Eq.~(\ref{con:atidleevolution}) can be rewritten according to the input field $c_{\rm in}(t)$ in a non-Markovian format as
\begin{equation} \label{con:atidleSolution}
\begin{aligned}
\dot{\tilde{a}}(t)&=  -i \omega_a \tilde{a} - \kappa_a \tilde{a} - i\zeta^*a  - 2i\chi \tilde{a}^{\dag}\tilde{a}^2 -2i\chi a^{\dag}a \tilde{a} \\
&-   \left [\sqrt{\tilde{g}_1}c_{\rm in} \left(  t -\frac{x}{c}  \right) + \sqrt{\tilde{g}_2} c_{\rm in} \left(t -   \frac{x-L}{c}  \right)\right ],
\end{aligned}
\end{equation}
which is similar to Eq.~(\ref{con:adynamics}). Obviously, when $\sqrt{g_1g_2} = \sqrt{\tilde{g}_1 \tilde{g}_2} =0$, the non-Markovian dynamics in Eqs.~(\ref{con:adynamics},\ref{con:atidleSolution}) reduces to be Markovian. 

Apart from non-Markovian interactions between the microring resonator and waveguide, the photon transport in Fig.~\ref{fig:system} is also influenced by the resonator's loss rate, and the coupling strengths among the Kerr material, clockwise and counterclockwise resonator modes. Based on these interactions concluded in Eq.~(\ref{eq:operatorDMaadd}), we can derive the non-Markovian input-output formalism as follows~\cite{gardiner1985input,fan2010input}. 

Take the interaction between waveguide and counterclockwise resonator mode as an example, the dynamics of the waveguide mode $b_{\omega}(\omega,t)$ can be solved by Eq.~(\ref{con:dkevolution}) as
\begin{equation} \label{con:domegatSolution}
\begin{aligned}
&b_{\omega}(\omega,t) = e^{-i\omega\left(t-t_0\right)}b_{\omega}(\omega,t_0) \\
& + \left (\sqrt{\frac{g_1}{2\pi}} e^{i\omega \frac{x}{c}} + \sqrt{\frac{g_2}{2\pi}} e^{i\omega \frac{x-L}{c}}\right )\int_{t_0}^t e^{-i\omega \left(t-t' \right)} a\left (t' \right)\mathrm{d}t'.
\end{aligned}
\end{equation}
By integrating Eq.~(\ref{con:dkevolution}) up to a final time $t_1>t$, the output field $b_{\rm out}(t)$ can be defined as the Fourier transform of $b_{\omega}\left(\omega,t_1\right)$~\cite{fan2010input,zhang2013non}, namely
\begin{equation} \label{con:boutCal}
\begin{aligned}
&~~~b_{\rm out}(t) = \frac{1}{\sqrt{2\pi}} \int_{-\infty}^{\infty} b_{\omega}(\omega,t_1) e^{-i\omega \left(t-t_1\right)}\mathrm{d}\omega\\
& = \frac{1}{\sqrt{2\pi}} \int_{-\infty}^{\infty}  e^{-i\omega\left(t-t_0\right)}b_{\omega}(\omega,t_0) \mathrm{d}\omega + \frac{1}{2\pi} \int_{t_0}^{t_1} \int_{-\infty}^{\infty}\\
&~~~  \left (\sqrt{g_1}  + \sqrt{g_2}  e^{-i \frac{\omega L}{c}}\right ) e^{-i\omega \left(t-t' - \frac{\omega x}{c} \right)}  \mathrm{d}\omega a\left (t' \right)\mathrm{d}t',
\end{aligned}
\end{equation}
where the first component equals to the input quantum field $b_{\rm in}(t)$ according to Eq.~(\ref{bindef}), and the second component represents delayed interactions analogous to  Eqs.~(\ref{con:adynamics},\ref{con:atidleSolution}). Then the overall input-output formalism for the system in Fig.~\ref{fig:system} can be represented as
\begin{small}
\begin{subequations} \label{eq:OneresonatorInputOutput}
\begin{numcases}{}
b_{\rm out}(t) = b_{\rm in}(t) +  \sqrt{g_1} a\left( t - \frac{x}{c}\right) +  \sqrt{g_2}  a\left(t - \frac{x-L}{c} \right), \label{boutIO}\\
c_{\rm out}(t) = c_{\rm in}(t) +  \sqrt{\tilde{g}_1} \tilde{a}\left( t + \frac{x}{c}\right) +  \sqrt{\tilde{g}_2}  \tilde{a}\left(t + \frac{x-L}{c} \right),\label{routIO}
\end{numcases}
\end{subequations}
\end{small}%
where $c_{\rm out}(t)$ is defined and derived in a similar approach according to  the Fourier transform of $c_{\omega}\left(\omega,t_1\right)$. Obviously, the input-output formalism in Eq.~(\ref{eq:OneresonatorInputOutput}) is non-Markovian due to the resonator state with a time delay $L/c$.

Based on the above input-output formalism in Eq.~(\ref{eq:OneresonatorInputOutput}), and the resonator dynamics in Eqs.~(\ref{con:adynamics},\ref{con:atidleSolution}), we firstly study the photon transport and blockade in Sec.~\ref{Sec:Scattering} without considering the interaction between the modes $a$ and $\tilde{a}$ by taking $\zeta =0$ for simplification, and the photon transport property can be evaluated by the scattering matrix for the output photons at the end of waveguide. Then we further consider the circumstance that $\zeta\neq 0$ in Sec.~\ref{Sec:PBmasterEquation}, where the photon transport property is evaluated by the eigenstates in the microring resonator.

\section{Photon transport evaluated by the scattering matrix} \label{Sec:Scattering}
The scattering matrix for $N$-photon transport can be defined as~\cite{fan2010input,SmatrixCal,xu2015input}
\begin{equation} \label{con:SmatrixDefine}
\begin{aligned}
S_{\nu_1\cdots \nu_N;\omega_1\cdots \omega_N} = \left\langle \nu_1\cdots \nu_N\right| S_N \left| \omega_1 \cdots \omega_N \right \rangle,
\end{aligned}
\end{equation}
where $\left| \omega_1 \cdots \omega_N \right \rangle$ represents the input photon states with the modes $\omega_1,\cdots,\omega_N$, $\left\langle \nu_1\cdots \nu_N\right|$ represents the output photon states with the modes $\nu_1,\cdots,\nu_N$, $S_N$ equals to the evolution operator $U_{\rm I}$ in the interaction picture from initial time $t_0 \rightarrow  -\infty$ to the terminal time $t_1 \rightarrow  \infty$ as
\begin{equation} \label{con:SDefine}
\begin{aligned}
S_N &= \lim_{t_0 \rightarrow -\infty,  t_1 \rightarrow \infty} U_{\rm I}\left(t_1,t_0\right)\\
& = \lim_{t_0 \rightarrow -\infty,  t_1 \rightarrow \infty} e^{iH_0 t_1} e^{-iH\left( t_1 - t_0 \right)}e^{iH_0 t_0},
\end{aligned}
\end{equation}
where  $H_0 = \omega_a \left(a^{\dag}a+\tilde{a}^{\dag}\tilde{a}\right)+ H_{\rm w}$ is the noninteracting free Hamiltonian in Eq.~(\ref{con:Htotal}).

We take the waveguide mode $b_{\omega}$ as an example. The operator for the input photon field $B_{\rm in}(\omega)$ can be defined in the frequency domain as
\begin{equation} \label{con:Ainomega} 
\begin{aligned}
B_{\rm in}(\omega) = e^{iHt_0}e^{-iH_0t_0} b_{\omega} e^{iH_0 t_0}e^{-iHt_0},
\end{aligned}
\end{equation}
and the output field operator  $B_{\rm out}(\omega)$ as
\begin{equation} \label{con:Aoutomega}
\begin{aligned}
B_{\rm out}(\omega) = e^{iHt_1}e^{-iH_0t_1} b_{\omega} e^{iH_0 t_1}e^{-iHt_1},
\end{aligned}
\end{equation}
where $b_{\omega}$ is that given by Eq.~(\ref{con:Hw}) and its relationship with the input field in the time domain has been clarified in Eq.~(\ref{eq:FreqOperator}).

The creation of the input and output scattering eigenstates can be represented as
\begin{subequations} \label{eq:CreatingInputOutput}
\begin{numcases}{}
B_{\rm in}^{\dag}(\omega) |0\rangle  = \left |\omega^+ \right\rangle,\\
B_{\rm out}^{\dag}(\nu) |0\rangle = \left |\nu^- \right\rangle ,
\end{numcases}
\end{subequations}
where $\left |\omega^+ \right\rangle$ represents the input photon field with the frequency $\omega$ and $\left |\nu^- \right\rangle$ represents the output photon field with the frequency $\nu$. Obviously, 
\begin{small}
\begin{equation} 
\begin{aligned}  \label{eq:AinOmegaCommut}
\left [ B_{\rm in}(\omega), B_{\rm in}^{\dag}(\nu)\right] =\left [ B_{\rm out}(\omega), B_{\rm out}^{\dag}(\nu)\right]  = \delta\left( \omega - \nu\right).
\end{aligned}
\end{equation}
\end{small}

\subsection{Single-photon transport}
When there is only one input photon in the waveguide, the dynamics will not be influenced by the coupling to the Kerr material because the nonlinear term $\chi a^{\dag}a^2$ in Eq.~(\ref{con:adynamics}) can always annihilate a single-photon. We take $\chi = h = 0$ in Eq.~(\ref{eq:operatorDMaadd}) for simplification. Then the scattering matrix $S_1$ for single-photon transport can be indexed by the input photon frequency $\omega$ and output photon frequency $\nu$ as
\begin{equation} \label{con:SDefineSinglePhoton}
\begin{aligned}
\left\langle \nu \right| S_1 \left|\omega \right\rangle = \left\langle 0 \right| B_{\rm out}(\nu) B_{\rm in}^{\dag}(\omega) \left| 0 \right\rangle.
\end{aligned}
\end{equation}

The input and output fields in Eq.~(\ref{boutIO}) can be rewritten based on Eqs.~(\ref{con:Ainomega},\ref{con:Aoutomega}) as
\begin{equation} \label{con:bintAinOmegaRelationCal2}
\begin{aligned}
b_{\rm in}(t) & = \frac{1}{\sqrt{2\pi}} \int_{-\infty}^{\infty} B_{\rm in}(\omega) e^{-i\omega t}\mathrm{d}\omega,
\end{aligned}
\end{equation}
and 
\begin{equation} \label{con:bouttAoutOmegaRelation}
\begin{aligned}
b_{\rm out}(t) &= \frac{1}{\sqrt{2\pi}} \int_{-\infty}^{\infty} B_{\rm out}(\omega) e^{-i\omega t}\mathrm{d}\omega.
\end{aligned}
\end{equation}

According to the detailed derivation in Appendix~\ref{Sec:AppendixSinglephoton},  the transport of single photon can be evaluated by the scattering matrix as
\begin{equation} \label{con:SSinglePhotonResultFinal}
\begin{aligned}
&\left\langle \nu \right| S_1 \left|\omega \right\rangle = t_{\omega}\delta\left(\omega-\nu \right),
\end{aligned}
\end{equation}
where
\begin{equation} \label{con:SinglePhotontw}
\begin{aligned}
 t_{\omega} = 1- \frac{g_1 + g_2 +2 \sqrt{g_1 g_2} \cos(\omega L/c) }{\kappa_a+i\left(\omega_a -\omega \right)}.
\end{aligned}
\end{equation}
Obviously, when $g_1 =g_2$ and $\cos(\omega L/c)= -1$, $\left\langle \nu \right| S_1 \left|\omega \right\rangle = \delta\left(\omega-\nu \right) $, and the transport of single photon in the waveguide will not be influenced by the resonator. When $g_1\neq g_2$, the single photon scattering matrix can be influenced by the coupling strength $g_1$, $g_2$, and the transmission delay $L/c$.

\begin{figure}[h]
\centerline{\includegraphics[width=1\columnwidth]{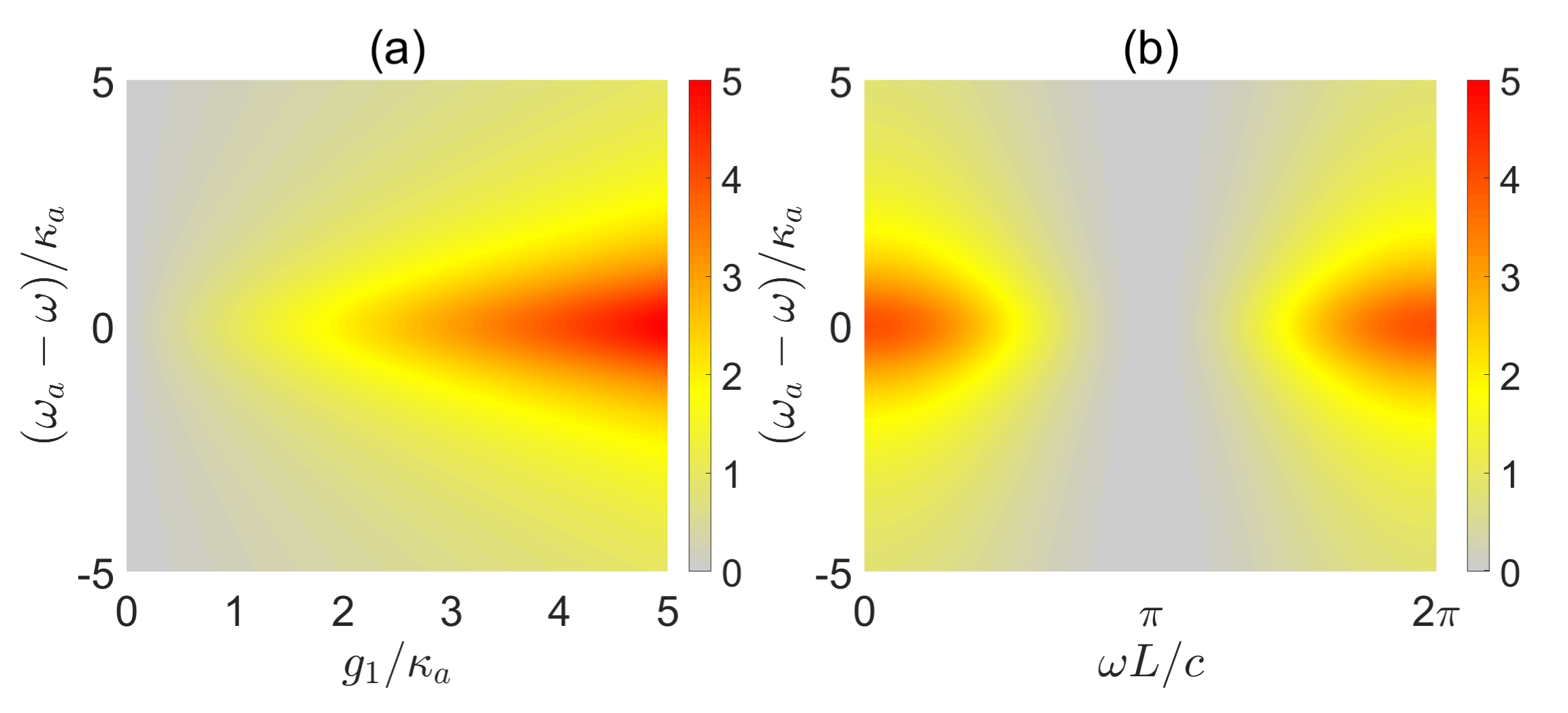}}
\caption{Single-photon transport evaluated by $\left| t_{\omega} -1\right|$ in Eq.~(\ref{con:SinglePhotontw}). In (a), $g_2/\kappa_a = 0$, the single-photon transport is influenced by the relative relationship among $g_1$, $\kappa_a$ and $\omega_a$. In (b), $g_1/\kappa_a =g_2/\kappa_a = 1$, the non-Markovian single-photon transport is influenced by $L$.}
	\label{fig:Kerr}
\end{figure}

To illustrate how the single-photon transport can be influenced by the non-Markovian interactions between the waveguide and microring resonator, we compare the amplitudes of the second component of the scattering matrix in Eq.~(\ref{con:SSinglePhotonResultFinal}) in Fig.~\ref{fig:Kerr}(a) for a Markovian case with  $g_1 \neq 0$, $g_2 = \tilde{g}_1 = \tilde{g}_2 = 0$, and in Fig.~\ref{fig:Kerr}(b) for a non-Markovian case with $g_1, g_2 \neq 0$, $\tilde{g}_1 = \tilde{g}_2 = 0$, respectively. Fig.~\ref{fig:Kerr}(a) shows that, in the Markovian interaction between the waveguide and resonator via single-point coupling, the transmission of single-photon is determined by the coupling strength between the waveguide and resonator. Fig.~\ref{fig:Kerr}(b) shows that, in the non-Markovian interaction realized by two-point coupling at $x$ and $x-L$, the single-photon transport is also influenced by the length of the waveguide between two interaction points $L$, and the photon can be ideally transported when $g_1= g_2$ and $\omega L/c = \pi$.

Both Eq.~(\ref{con:SinglePhotontw}) and the numerical simulations in Fig.~\ref{fig:Kerr} show that, the transport of a single photon can be modulated by the phase difference during its transmission between the two coupled points in the waveguide, namely the parameter $\omega L/c$ in Eq.~(\ref{con:SinglePhotontw}), and this is similar to the classical circumstance that electric fields transports in a waveguide coupled to a mircoring resonator at two points~\cite{shoman2020measuring}.

\subsection{Two-photon transport and blockade} \label{TwophotonTransporation}
When there are two input photons, the two-photon transport can be affected by the Kerr nonlinearity due to the Hamiltonian in Eq.~(\ref{con:Hkerr}), which is different from the single-photon circumstance~\cite{SmatrixCal}. Generalized from the above single-photon case, in the two-photon transport, we denote the input photon frequencies as $\omega_1$ and $\omega_2$, and the output photon frequencies as $\nu_1$ and $\nu_2$. Thus by taking $N=2$ in Eq.~(\ref{con:SmatrixDefine}), the two-photon scattering matrix reads~\cite{fan2010input}
\begin{equation} \label{con:TwoPhotonMatrix}
\begin{aligned}
&\left\langle \nu_1 \nu_2 \right| S_2 \left|\omega_1\omega_2 \right\rangle \\
=& \left\langle 0 \right| B_{\rm out}\left(\nu_1\right) B_{\rm out}\left(\nu_2\right) B_{\rm in}^{\dag}\left(\omega_1\right) B_{\rm in}^{\dag}\left(\omega_2\right) \left| 0 \right\rangle.
\end{aligned}
\end{equation}

Then the scattering matrix for two-photon transport can be further derived as
\begin{widetext}
\begin{equation} \label{con:TwoPhotonMatrixResultGeneral}
\begin{aligned}
&\left\langle \nu_1 \nu_2 \right| S_2 \left|\omega_1\omega_2 \right\rangle =t_{\nu_1}t_{\nu_2} \left[ \delta\left( \nu_1 - \omega_1\right) \delta \left( \nu_2 - \omega_2\right)  +   \delta\left( \nu_2 - \omega_1\right) \delta\left( \nu_1 - \omega_2\right) \right]\\
&+4i\left( t_{\nu_1} \eta_{\nu_2} +  t_{\nu_2} \eta_{\nu_1}\right) \frac{\chi\kappa_a \left[2\kappa_a +i \left(2\omega_a - \nu_1- \nu_2 \right) \right]\Gamma^*\left(\nu_1\right)\Gamma^*\left(\nu_2\right) \Gamma\left(\omega_1\right) \Gamma\left(\omega_2\right)  }{\kappa_a \left[2\kappa_a +i \left(2\omega_a - \nu_1 - \nu_2 \right) \right]+2\pi i\chi   \left(g_1 + g_2 \right) }\delta \left( \nu_1+\nu_2-\omega_1- \omega_2\right),
\end{aligned}
\end{equation}
\end{widetext}
where $t_{\nu_j}$ with $j =1,2$ represents the single-photon process by taking $\omega = \nu_j$ in Eq.~(\ref{con:SinglePhotontw}), namely
\begin{equation} \label{con:EtaDefine}
\begin{aligned}
\eta_{\omega}=1- i\frac{2 \left(\omega_a-\omega\right) }{\kappa_a+i \left(\omega_a-\omega\right) },
\end{aligned}
\end{equation}
which has a similar format as $t_{\omega}$. We additionally define the following function for the input and output photon frequencies
\begin{equation} \label{con:GammaDefine}
\begin{aligned}
\Gamma(\omega)=\frac{\sqrt{g_1} + \sqrt{g_2} e^{i\omega \frac{L}{c}}}{\kappa_a+i \left(\omega_a-\omega\right)  }.
\end{aligned}
\end{equation}
More details on $\eta_{\omega}$ and $\Gamma(\omega)$ are given in Appendix~\ref{Sec:AppendixTwophoton}. The scattering matrix in Eq.~(\ref{con:TwoPhotonMatrixResultGeneral}) shows that, in this non-Markovian network with Kerr nonlinearity, the two-photon transport is not simply a combination of two single-photon transport processes, namely the components in the first line of the right hand side of Eq.~(\ref{con:TwoPhotonMatrixResultGeneral}). A more complex term is the correlated term in the second line of Eq.~(\ref{con:TwoPhotonMatrixResultGeneral}), and this is induced by the nonlinear interactions between the microring resonator and the Kerr material, which makes it different from the transport of classical fields.

On the other hand, the photon blockade effect can be evaluated by the second order correlation functions for the field at the output end of the waveguide~\cite{TwoSMwaveguideTwoCav,wang2022few}.
When there are two input photons from the same direction, the average number of photons collected at the output end of the waveguide reads~\cite{TwoSMwaveguideTwoCav}
\begin{equation} \label{con:AveragePhotonOutput}
\begin{aligned}
N_{\rm out}(t) &= \left \langle b_{\rm out}^{\dag}(t)b_{\rm out}(t) \right\rangle,
\end{aligned}
\end{equation}
where $b_{\rm out}(t)$ is given by Eq.~\eqref{eq:OneresonatorInputOutput}. Then the second-order correlation function of the output field can be evaluated between two time points $t$ and $t'$ as~\cite{wang2022few,TwoSMwaveguideTwoCav,liew2010single}
\begin{equation} \label{con:g2tauttp}
\begin{aligned}
g_{\rm out}^{(2)}(t-t') = \frac{\left \langle b_{\rm out}^{\dag}(t')b_{\rm out}^{\dag}(t)b_{\rm out}(t)b_{\rm out}(t') \right\rangle}{N_{\rm out}(t)N_{\rm out}(t') }.
\end{aligned}
\end{equation} 
In the following, we study how the photon blockade evaluated by the above correlation function can be influenced by the non-Markovian interactions between the waveguide and microring resonator.

When there are two input photons with frequencies $\omega_1$ and $\omega_2$, we denote the input photon state as $\left|\psi_{\rm in} \right\rangle $ as a product state of two coherent states with the identical amplitude $\alpha$ in the format of a coherent state as~\cite{wang2022few}
\begin{equation} \label{con:g2tauPart1}
\begin{aligned}
\left|\psi_{\rm in} \right\rangle = \left|\alpha_1 \right\rangle \otimes  \left|\alpha_2 \right\rangle,
\end{aligned}
\end{equation}
where $\left|\alpha_j \right\rangle = |0 \rangle + \alpha |1\rangle_j + \left(\alpha^2/\sqrt{2}\right) |2\rangle_j + O\left(\alpha^3\right)$ with the index $j =1,2$ representing the two input photons. Then combined with the format of $b_{\rm out}(t)$ in Eq.~(\ref{con:bouttAoutOmegaRelation}), one of the denominator in Eq.~(\ref{con:g2tauttp}) for an input photon mode $\omega$ reads~\cite{wang2022few}
\begin{equation} \label{con:g2tauPart1}
\begin{aligned}
 &\left \langle \psi_{\rm in} \right| b_{\rm out}^{\dag}(t)b_{\rm out}(t) \left|\psi_{\rm in}\right\rangle\\
 =& |\alpha|^2 \int_{-\infty}^{\infty} \left \langle \nu \right| e^{i\nu t}\mathrm{d}\nu  B_{\rm out}^{\dag}(\omega)  |0\rangle \\
 &\langle 0| \int_{-\infty}^{\infty} B_{\rm out}(\omega') e^{-i\nu' t}\left | \nu' \right \rangle \mathrm{d}\nu'\\
=& |\alpha|^2 \left| t_{\omega}\right|^2,
\end{aligned}
\end{equation} 
where $t_{\omega}$ is given by Eq.~(\ref{con:SinglePhotontw}). Besides, for the numerator in Eq.~(\ref{con:g2tauttp}) with the diverse $\tau$ between $t$ and $t'$, 
\begin{equation} \label{con:g2tauUpper}
\begin{aligned}
 &\left \langle \psi_{\rm in} \right| b_{\rm out}^{\dag}(t)b_{\rm out}^{\dag}(t+\tau)b_{\rm out}(t+\tau)b_{\rm out}(t) \left|\psi_{\rm in}\right\rangle\\
=&|\alpha|^4  \iint_{-\infty}^{\infty}   S_2^* \left( \nu_1,\nu_2;  \omega_1,\omega_2\right) e^{i\nu_1t} e^{i\nu_2(t+\tau)}\mathrm{d}\nu_1 \mathrm{d}\nu_2\\
&\times \iint_{-\infty}^{\infty}   S_2 \left( \nu_1,\nu_2;  \omega_1,\omega_2\right) e^{-i\nu_1t} e^{-i\nu_2(t+\tau)}\mathrm{d}\nu_1 \mathrm{d}\nu_2\\
\triangleq & |\alpha|^4 \left|t_{\omega_1}t_{\omega_2} + T\left(\omega_1,\omega_2 \right)   \right|^2,
\end{aligned}
\end{equation} 
where the elements of the scattering matrix $S_2 \left(\nu_1,\nu_2;  \omega_1,\omega_2\right)=\left\langle \nu_1 \nu_2 \right| S_2 \left|\omega_1\omega_2 \right\rangle$ have been derived in Eq.~(\ref{con:TwoPhotonMatrixResultGeneral}), and $T\left(\omega_1,\omega_2 \right)$ is influenced by the Kerr nonlinearity and non-Markovian coupling between the microring resonator and waveguide.

Then the second order correlation function $g_{\rm out}^{(2)}(0)$ for two-photon transport can be derived according to the scattering matrix as
\begin{equation} \label{con:g2out0}
\begin{aligned}
 &g_{\rm out}^{(2)}(0) = \frac{\left|t_{\omega_1}t_{\omega_2} + T\left(\omega_1,\omega_2 \right)   \right|^2}{\left| t_{\omega_1}\right|^2 \left| t_{\omega_2}\right|^2},
\end{aligned}
\end{equation} 
and the photon blockade occurs due to the cancellation by $T\left(\omega_1,\omega_2 \right)$ upon $t_{\omega_1}t_{\omega_2}$~\cite{wang2022few}, resulting in $g_{\rm out}^{(2)}(0)<1$.

\begin{figure}[h]
\centerline{\includegraphics[width=1\columnwidth]{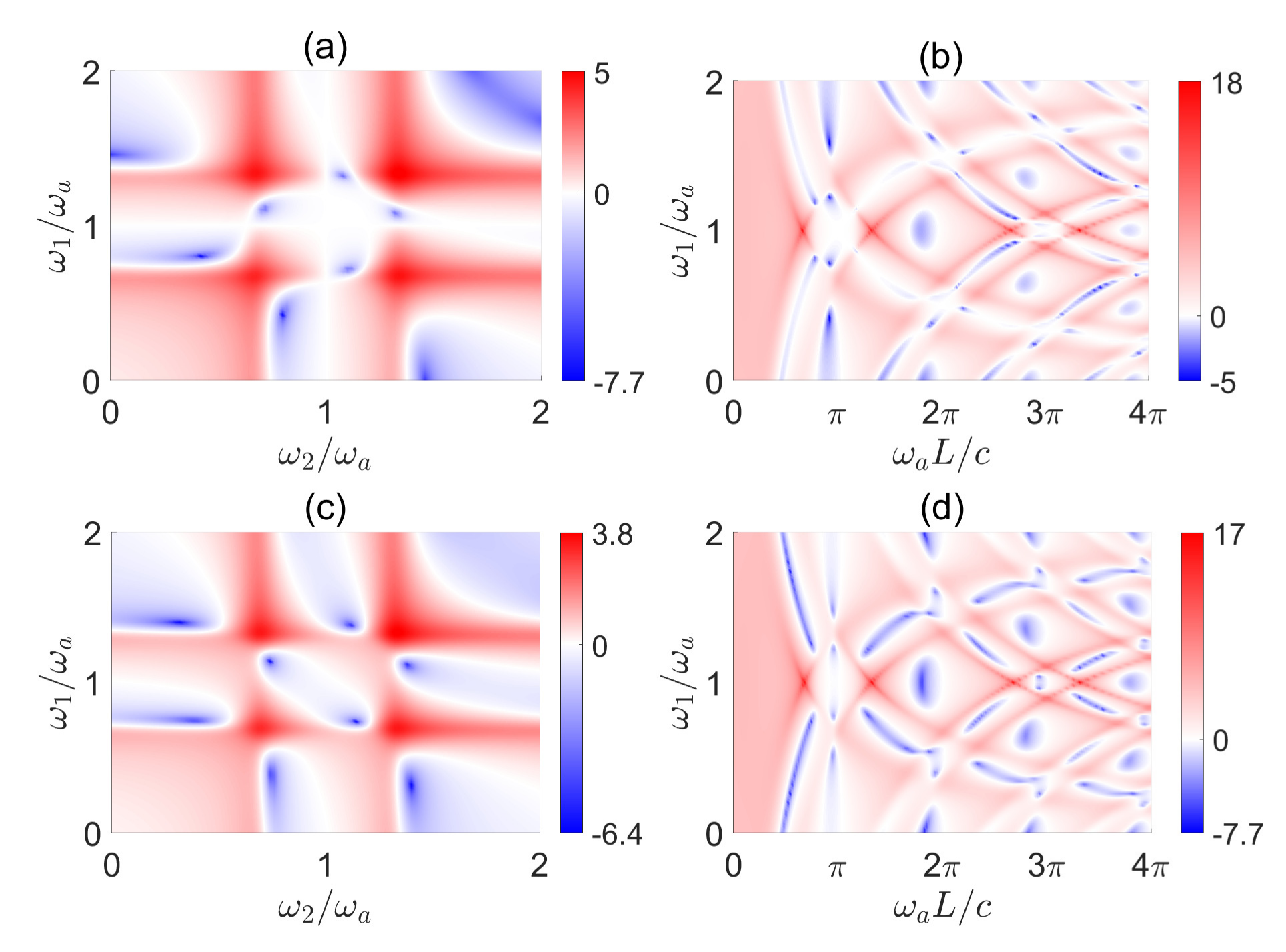}}
\caption{Second order correlation function $\ln \left[ g_{\rm out}^{(2)}(0)\right]$ of the output field is influenced by input photon frequencies and other parameters when $\kappa_a = 1$GHz and $\chi = 0.1$GHz. In (a) and (b), $g_1 = g_2 = 1$GHz. In (c) and (d), $g_2 = 2g_1 = 4/3$GHz. In (a) and (c), $\omega_a L/c = \pi$. In (b) and (d), $\omega_1 + \omega_2 = 2\omega_a$.}
	\label{fig:g2Scatter}
\end{figure}
As compared in Fig.~\ref{fig:g2Scatter}, the occurrence of photon blockade is influenced by the coupling strength between the resonator and waveguide, the frequencies of two input photons, and the non-Markovian interaction evaluated by the length of the waveguide $L$ between two coupled points. For example, the comparison between Figs.~\ref{fig:g2Scatter}(a,c) for two input photons shows that the values of $g_1$ and $g_2$ can influence the occurrence of photon blockade.  Due to the format of scattering matrix in the second line of Eq.~(\ref{con:TwoPhotonMatrixResultGeneral}), we take $\omega_1 + \omega_2 = 2\omega_a$ in the simulations of Figs.~\ref{fig:g2Scatter}(b,d). In this setting, the second order correlation functions of the output field can be influenced by both $L$ and the frequency of each input photon. The simulations in Fig.~\ref{fig:g2Scatter}(d) also show that, the diverse between $g_1$ and $g_2$ can induce blockade effect when the frequencies of two input photons are different. 

The above comparisons in Fig.~\ref{fig:g2Scatter} indicate that, photon blockade is influenced by the two-point coupling strengths between resonator and waveguide, the input photon frequencies and the length of waveguide between two coupled points. As determined by the scattering matrix in Eq.~(\ref{con:TwoPhotonMatrixResultGeneral}), because of the existence of Kerr nonlinearity $\chi$ and decaying rate to the environment $\kappa_a$, the photon blockade will not only occur when $\omega_aL/c = \mathbf{n}\pi$ with $\mathbf{n} = 1,2,\cdots$, but also occur when $\omega_aL/c \neq \mathbf{n}\pi$, and the input photon frequencies do not equal $\omega_a$. This is supported by the numerical simulations in Figs.~\ref{fig:g2Scatter}(a,c) that  $\ln \left[ g_{\rm out}^{(2)}(0)\right]<0$ when $\omega_1 = \omega_2 \neq \omega_a$, and especially in Fig.~\ref{fig:g2Scatter}(d) that $\ln \left[ g_{\rm out}^{(2)}(0)\right]<0$ when $\omega_aL/c \neq \mathbf{n}\pi$ and $\omega_1, \omega_2 \neq \omega_a$.

\section{Non-Markovian photon blockade with a driven resonator} \label{Sec:PBmasterEquation}
Different from Sec.~\ref{Sec:Scattering} where the microring resonator is driven by input photons transmitted through waveguide in the format of quantum fields with continuous modes, in this section, we study the photon blockade effect when the microring resonator is driven by classical fields. Above all, the non-Markovian interaction between the waveguide and microring resonator can be equivalently represented with the master equation by  tracing out the waveguide modes~\cite{pichler2015quantum,WallraffWaveguide}. Within the Markovian approximation by regarding the delayed components such as in Eqs.~(\ref{con:adynamics},\ref{con:atidleSolution},\ref{eq:OneresonatorInputOutput}) as delayed phase related with the resonator frequency $\omega_a$~\cite{WallraffWaveguide}, i.e.,
\begin{subequations} \label{con:delayapproximation}
\begin{numcases}{}
a\left(t+\frac{L}{c}\right) \approx a(t)e^{-i\omega_a L/c},\\
a^{\dag}\left(t+\frac{L}{c}\right) \approx a^{\dag}(t)e^{i\omega_a L/c},
\end{numcases}
\end{subequations}
the master equation for the non-Markovian interaction between microring resonator and waveguide in Fig.~\ref{fig:system} can be represented as~\cite{LPB2023tunable,lu2023dressed,kockum2018decoherence,soro2022chiral}
\begin{small}
\begin{equation} \label{con:masterequation}
\begin{aligned}
\dot{\rho}(t) =  &-i \left[ H_{a,\tilde{a}} + \sin \left(\frac{\omega_a L}{c} \right) \left(\sqrt{g_1g_2}   a^{\dag}a + \sqrt{\tilde{g}_1\tilde{g}_2} \tilde{a}^{\dag}\tilde{a}\right), \rho\right] \\
&+  \left [\frac{g_1 +g_2}{2} +   \sqrt{g_1 g_2} \cos \left(\frac{\omega_a L}{c} \right)  + \kappa_a  \right] \mathcal{L}_{a}[\rho] \\
& +  \left [\frac{\tilde{g}_1 +\tilde{g}_2}{2} +   \sqrt{\tilde{g}_1\tilde{g}_2} \cos \left(\frac{\omega_a L}{c} \right)  + \kappa_a  \right] \mathcal{L}_{\tilde{a}}[\rho],
\end{aligned}
\end{equation} 
\end{small}%
where $H_{a,\tilde{a}}$ represents the free Hamiltonian of the microring resonator independent from its interaction with the waveguide, and $\mathcal{L}_{O}[\rho] =  2O \rho O^{\dag} -\rho O^{\dag} O -  O^{\dag} O \rho$ for an arbitrary operator $O$. The master equation~(\ref{con:masterequation}) can be similarly derived via the approach in Refs.~\cite{pichler2015quantum,ClerkPRX} or by replacing the atomic operator in Refs.~\cite{kockum2018decoherence,soro2022chiral} with resonator operators, and more details are omitted.

Based on Eq.~(\ref{con:masterequation}), we denote~\cite{Jinghui2024chiral}
\begin{subequations} \label{con:geffDef}
\begin{numcases}{}
g_{\rm eff}^-  = \frac{g_1 +g_2}{2} +   \sqrt{g_1 g_2} \cos \left(\frac{\omega_a L}{c}\right)  + \kappa_a \geq 0,\\
g_{\rm eff}^+ = \frac{\tilde{g}_1 +\tilde{g}_2}{2} +   \sqrt{\tilde{g}_1  \tilde{g}_2} \cos \left(\frac{\omega_a L}{c}\right)  + \kappa_a \geq 0,
\end{numcases}
\end{subequations}
and 
\begin{subequations} \label{con:geffDefFirstHalf}
\begin{numcases}{}
\hat{g}_{\rm eff}^-  =  \sqrt{g_1g_2} \sin \left(\frac{\omega_a L}{c} \right),\\
\hat{g}_{\rm eff}^+ = \sqrt{\tilde{g}_1\tilde{g}_2} \sin \left(\frac{\omega_a L}{c} \right),
\end{numcases}
\end{subequations}
then the effective Hamiltonian can be represented as~\cite{Jinghui2024chiral}
\begin{equation} \label{con:HeffV2}
\begin{aligned}
H_{\rm eff} =& \zeta a^{\dag} \tilde{a} + \zeta^*a \tilde{a}^{\dag}  \\
& + \chi\left( a^{\dag}a^{\dag}aa +  \tilde{a}^{\dag}\tilde{a}^{\dag}\tilde{a}\tilde{a}+ 2a^{\dag}a\tilde{a}^{\dag}\tilde{a}\right)\\
 &+ \left( \hat{g}_{\rm eff}^- - ig_{\rm eff}^- \right)a^{\dag}a + \left( \hat{g}_{\rm eff}^+ - i g_{\rm eff}^+ \right)\tilde{a}^{\dag}\tilde{a}.
\end{aligned}
\end{equation}
Further combined with the external resonant drive~\cite{LPB2023tunable}
\begin{equation} \label{con:Hdformat}
\begin{aligned}
H_{\rm d} =\epsilon \left(a + a^{\dag} \right),
\end{aligned}
\end{equation}
the total Hamiltonian of the system can be represented as $H_{\rm tot} = H_{\rm eff} + H_{\rm d}$. We assume that $\epsilon$ is bounded such that there are at most two photons in the resonator. Thus the photonic states in the microring resonator with clockwise mode (indexed by the operator $\tilde{a}$) and counterclockwise mode (indexed by the operator $a$) can be represented as~\cite{liu2023parametric,Jinghui2024chiral}  
\begin{equation} \label{con:ResonatorPhotonState}
\begin{aligned}
|\psi \rangle =\sum_{m,n} C_{mn} |m,n\rangle = \sum_{m,n} C_{mn} |m\rangle \otimes |n\rangle,
\end{aligned}
\end{equation} 
where $m$ represents the number of photons of the counterclockwise mode, $n$ represents the number of photons of the clockwise mode, and there are at most two photons in the coupled system such that $m+n \leq 2$.

The dynamics of the state vector in Eq.~(\ref{con:ResonatorPhotonState}) is governed by the Schr\"{o}dinger equation $i \hbar  |\dot{\psi}(t) \rangle = H_{\rm tot} |\psi (t) \rangle$, then the evolution of the amplitudes for states in Eq.~(\ref{con:ResonatorPhotonState}) reads~\cite{TwoSMwaveguideTwoCav}
\begin{subequations} \label{con:AmplitudeEvolution}
\begin{numcases}{}
\dot{C}_{00} = -i \epsilon C_{10} ,\\
\dot{C}_{10} = -i \epsilon C_{00} -i\sqrt{2}\epsilon C_{20}-i\zeta C_{01} \notag\\
~~~~~~~~ -  \left(g_{\rm eff}^- +i \hat{g}_{\rm eff}^-  \right) C_{10}  ,\\
\dot{C}_{01} = -i \epsilon C_{11}  -i\zeta^* C_{10}  -  \left(g_{\rm eff}^+ +i \hat{g}_{\rm eff}^+  \right) C_{01} , \\
\dot{C}_{11} = -i \epsilon C_{01}- i\sqrt{2}\zeta C_{02} - i\sqrt{2}\zeta^* C_{20}  \notag\\
~~~~~~~~ - \left(g_{\rm eff}^- + g_{\rm eff}^+ + i \hat{g}_{\rm eff}^- + i\hat{g}_{\rm eff}^+ + 2i \chi \right)C_{11},\\
\dot{C}_{02} = - i \sqrt{2}\zeta^* C_{11} -  2\left(g_{\rm eff}^+ +i \hat{g}_{\rm eff}^+ +i\chi \right)C_{02} ,\\
\dot{C}_{20} = -i\sqrt{2}\zeta C_{11}  -i\sqrt{2} \epsilon  C_{10} \notag\\
~~~~~~~~ - 2 \left(g_{\rm eff}^- +i \hat{g}_{\rm eff}^- +i\chi \right)C_{20},
\end{numcases}
\end{subequations}
where we adopt the relationship between the resonator operators and their eigenstates as $a|m\rangle = \sqrt{m}|m-1\rangle $ when $m>0$, $a^{\dag}|m\rangle = \sqrt{m+1}|m+1\rangle $, $a^{\dag}a|m\rangle = m |m\rangle$, $\tilde{a}|n\rangle = \sqrt{n}|n-1\rangle $ when $n>0$, $\tilde{a}^{\dag}|n\rangle = \sqrt{n+1}|n+1\rangle $ and $\tilde{a}^{\dag}\tilde{a}|n\rangle = n |n\rangle$.

Due to the input-output formalism in Eq.~(\ref{eq:OneresonatorInputOutput}) without considering the input quantum fields such as $b_{\rm in}(t)$ and $c_{\rm in}(t)$, we consider the circumstance that the microring resonator is only driven by classical fields modeled in the Hamiltonian $H_{\rm d}$ in Eq.~(\ref{con:Hdformat}), then the correlations in the output field are only determined by the correlations of intracavity fields, similar to the circumstances in Refs.~\cite{TwoSMwaveguideTwoCav,liu2023parametric}. Considering that the mean value of photon number in the resonator satisfies that
\begin{equation} \label{con:adaMeanvalue}
\begin{aligned}
&\left \langle a^{\dag}a\right\rangle = \mathrm{Tr} \left( a^{\dag}a  |\psi \rangle \langle \psi | \right)\\
=&\mathrm{Tr} \left( a^{\dag}a  \sum_{m,n} C_{mn} |m,n\rangle  \sum_{m',n'} C_{m'n'}^*  \langle m',n' | \right)\\
=& \sum_{m,n} \left|C_{mn}\right|^2 \left( a^{\dag}a   |m,n\rangle   \langle m,n | \right)\\
=& \sum_{m,n} m \left|C_{mn}\right|^2,
\end{aligned}
\end{equation}
and further according to the calculation and simplification methods in Refs.~\cite{liew2010single,liu2023parametric},  
\begin{equation} \label{con:g2simplification}
\begin{aligned}
\left\langle a^{\dag}a^{\dag}aa \right\rangle = \sum_{m,n} m (m-1)\left|C_{mn}\right|^2.
\end{aligned}
\end{equation}
As a result, the second order correlation function when  $t= t'$ in Eq.~(\ref{con:g2tauttp}) can be evaluated by the intracavity states and approximated as~\cite{liu2023parametric}
\begin{small}
\begin{equation} \label{con:g20Solution}
\begin{aligned}
g_{\rm out}^{(2)}(0) &= \frac{\left \langle b_{\rm out}^{\dag}(t)b_{\rm out}^{\dag}(t)b_{\rm out}(t)b_{\rm out}(t) \right\rangle}{N_{\rm out}(t)N_{\rm out}(t) }\\
&\approx  \frac{ \sum_{m,n} m (m-1)\left|C_{mn}\right|^2}{\left(\sum_{m',n'} m' \left|C_{m'n'}\right|^2\right)^2}\\
&=\frac{2\left|C_{20}\right|^2}{\left(2\left|C_{20}\right|^2+ \left|C_{10}\right|^2 + \left|C_{11}\right|^2\right)^2},
\end{aligned}
\end{equation} 
\end{small}%
which is independent from $C_{00}$.

\begin{figure}[h]
\centerline{\includegraphics[width=1\columnwidth]{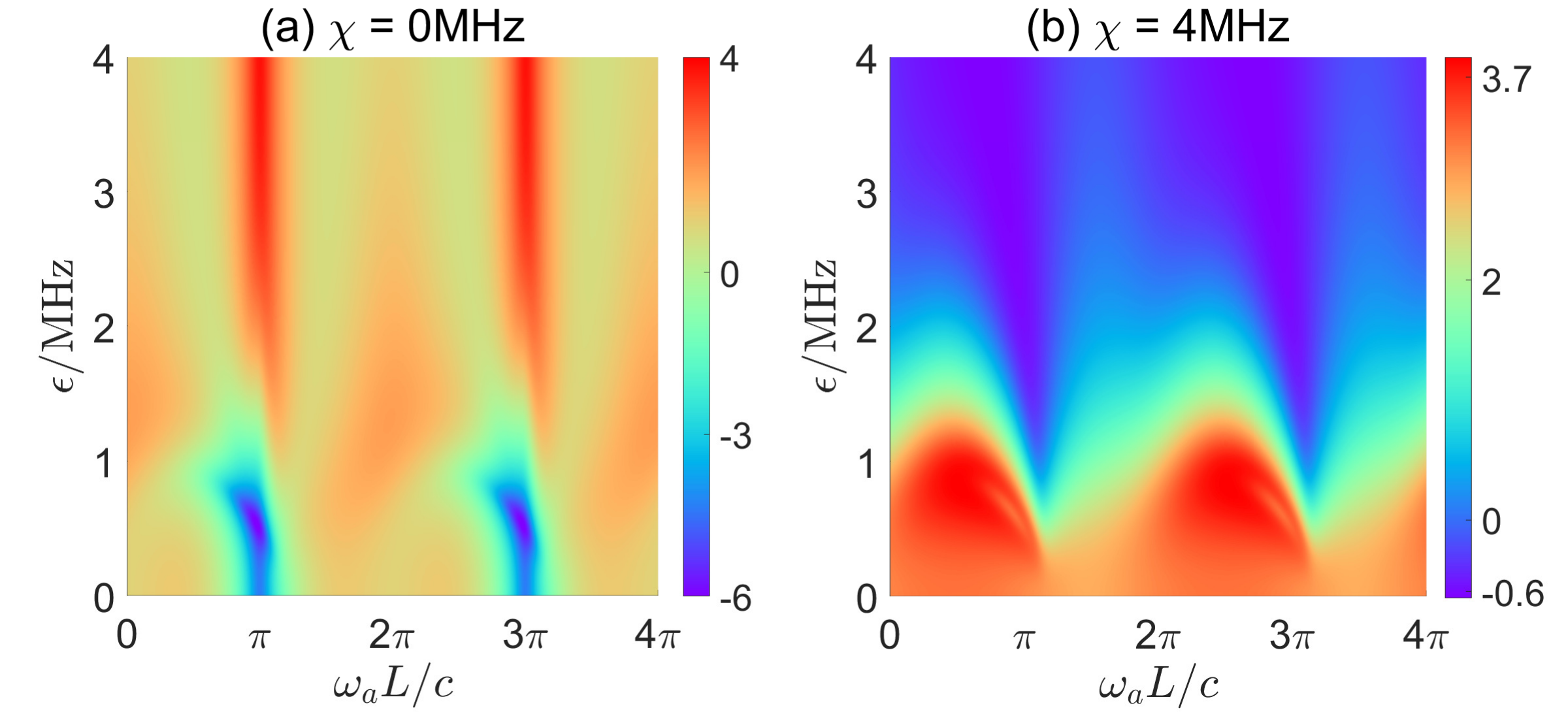}}
\caption{Second order correlation function $\ln \left[ g_{\rm out}^{(2)}(0)\right]$ influenced by Kerr nonlinearity when $g_0 = 1$GHz, $\kappa_a = 0.2$GHz and $\zeta = 0.5$GHz.}
	\label{fig:g2simulation}
\end{figure}
To clarify the photon blockade effect influenced by the Kerr nonlinearity, we take $g_1 = g_2 = \tilde{g}_1 = \tilde{g}_2 = g_0$ and $\zeta = \zeta^*$, then the steady values of the amplitudes for the photonic states in the microring resonator can be solved when $\dot{C}_{mn}$ at the lefthand side of Eq.~(\ref{con:AmplitudeEvolution}) equals zero. Then the second order correlations at the waveguide output end directly coupled with the resonator mode $a$ can be evaluated by Eq.~(\ref{con:g20Solution}), and simulated as Fig.~\ref{fig:g2simulation}. The comparison between Fig.~\ref{fig:g2simulation}(a) and Fig.~\ref{fig:g2simulation}(b) illustrates that the non-Markovian interaction between the waveguide and microring resonator can induce blockade effects. As in Fig.~\ref{fig:g2simulation} (a), when the microring resonator is not coupled to the Kerr material, $\ln \left[ g_{\rm out}^{(2)}(0)\right] < 0$  only when $\epsilon$ is small and $\omega_aL/c = (2 \mathbf{n} +1)\pi$ with $\mathbf{n} = 0, 1,\cdots$. However, as simulated in Fig.~\ref{fig:g2simulation}(b), when the microring resonator is coupled to the Kerr material, the nonlinear terms can annihilate two-photon states, and the photon blockade can occur when the drive field applied upon the resonator is stronger.

\begin{figure}[h]
\centerline{\includegraphics[width=1\columnwidth]{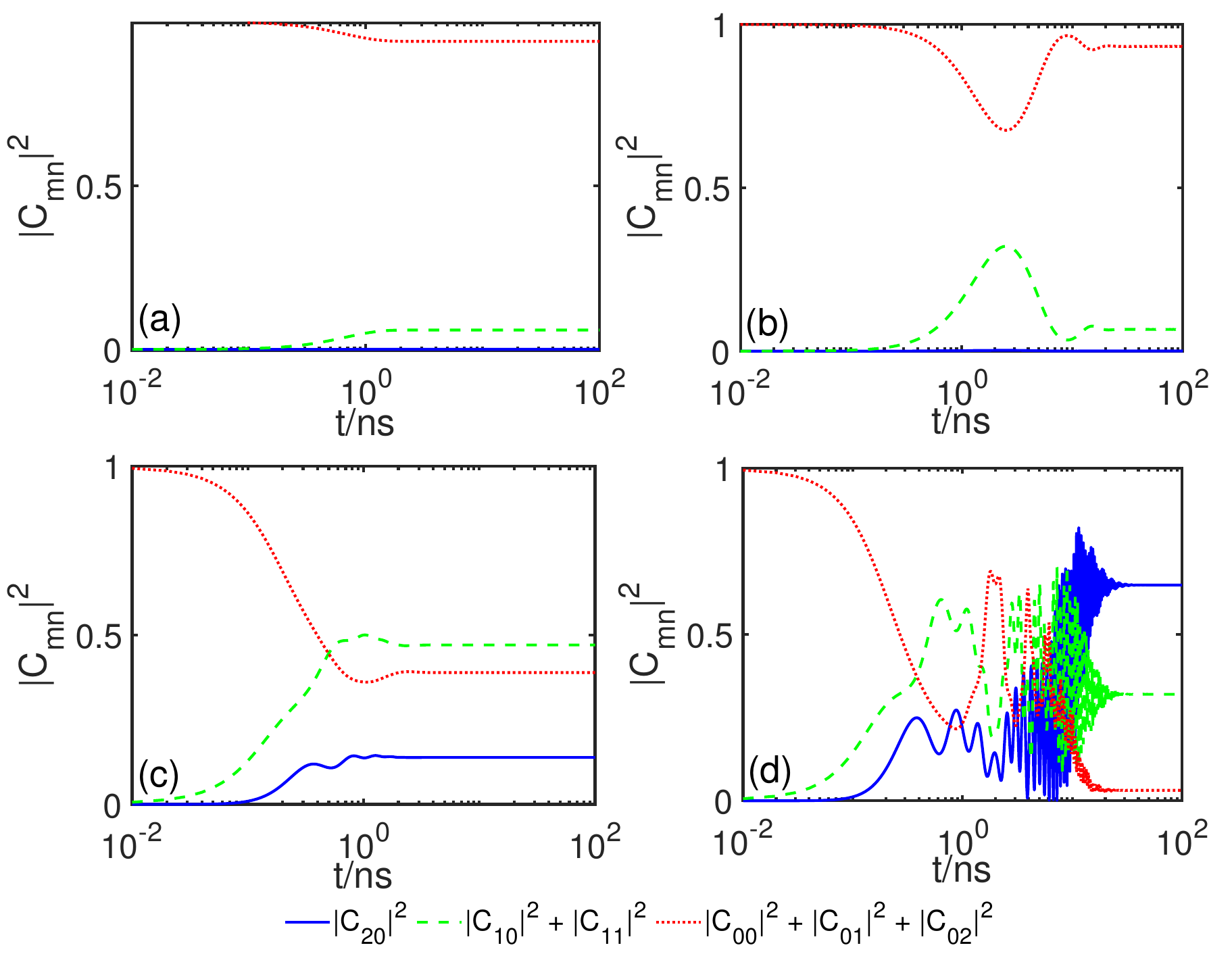}}
\caption{The photon states in the resonator with $g_0 = 1$GHz, $\chi = 4$GHz, $\kappa_a = 0.2$GHz and $\zeta = 0.5$GHz. As for the non-Markoivan parameter settings, $\omega_aL/c = \pi/3$ in (a) and (c), while $\omega_aL/c = \pi$ in (b) and (d). As for the classical driving field, $\epsilon  = 0.5$GHz in (a) and (b), while $\epsilon  = 4$GHz in (c) and (d). }
	\label{fig:PhotonPopulation}
\end{figure}

To further clarify the photon number states in the microring resonator, the amplitudes of quantum states are simulated in Fig.~\ref{fig:PhotonPopulation}. When the drive field is weak as $\epsilon = 1$GHz, as in Fig.~\ref{fig:PhotonPopulation}(a) with $\omega_a L/c = \pi/3$, the effective coupling between the waveguide and resonator is nonzero, which means that the photons in the resonator can be transmitted into the waveguide. However, because the driving field is weak and $\kappa_a>0$, there are almost no two-photon states in the resonator. When $\omega_a L = \pi$ and $\epsilon = 1$GHz in Fig.~\ref{fig:PhotonPopulation}(b), the effective coupling between the waveguide and resonator is zero according to Eq.~(\ref{con:geffDef}), then the drive applied upon the resonator can compensate its decaying to the environment, then there can be two-photon states in the resonator. Based on this, when the drive field is stronger, as in Figs.~\ref{fig:PhotonPopulation}(c,d) with $\epsilon = 4$GHz, the populations for two-photon states can be larger, and photon blockade can occur due to the area in Fig.~\ref{fig:g2simulation}(b) with  $\ln \left[ g_{\rm out}^{(2)}(0)\right] < 0$. For example, according to the steady numerical results in Fig.~\ref{fig:PhotonPopulation}(c),  $g_{\rm out}^{(2)}(0) \approx 0.4948<1$, and in Fig.~\ref{fig:PhotonPopulation}(d), $g_{\rm out}^{(2)}(0) \approx 0.4961<1$.  Besides, the comparison between Figs.~\ref{fig:PhotonPopulation}(c,d) also illustrates that the non-Markovian interaction between resonator and waveguide evaluated by $\omega_a L$ can influence the steady photonic states in the microring resonator. When $\omega_aL/c = \pi$ in Fig.~\ref{fig:PhotonPopulation}(d), the counterclockwise photon mode cannot be easily decayed to the waveguide, then the population for the steady two-photon states can be larger than that in Fig.~\ref{fig:PhotonPopulation}(c).

As for more details on the experimental realizations of the proposal in this paper, according to the on-chip fabrications in Refs.~\cite{chen2024ultralow,kazakov2024active,bogaerts2012silicon,churaev2023heterogeneously}, the resonant frequency of the microring resonator can be 10 GHz to 100 GHz, the length of the waveguide constructing the coherent feedback loop can be 0.1 mm to 10 mm~\cite{chen2024ultralow}, then the non-Markovian proposal realized by microring resonators coupled to the waveguide at two distinct points can be fabricated on-chip with the length of several millimeters~\cite{chen2024ultralow,silverstone2015qubit,wang2020integrated,elshaari2020hybrid,pelucchi2022potential,bao2023very,weaver2024integrated}.

Based on the above parameter settings of microring resonators, the proposal in this paper can be further generalized to complex networks with multiple resonators. For example, a one-dimensional array of microring resonators can be coupled to a waveguide~\cite{tang2022nonreciprocal,hsu2023chip,chen2024ultralow,pan2023parallel}, and the non-Markovian properties in such networks not only arise from the two-point coupling between one microring resonator and the waveguide as in this paper, but also arises from the transmission delays of photons between two microring resonators. 
Additionally, two-dimensional networks in the format of topological microring lattices~\cite{hafezi2011robust,song2025observation,gao2023topological,lin2023high} can be used to construct coupled
resonator optical waveguides. On one hand, in the one-dimensional and two-dimensional microring resonator networks, the edge states can be topologically-protected such that they are
immune to defect scattering~\cite{afzal2021topological}. On the other hand, by coupling the topological microring resonator array to a waveguide, the topological property of the topological system (i.e., band structure, nonreciprocal property, and so on) can be detected and evaluated via the output end of the waveguide~\cite{tang2022nonreciprocal,li2025nonreciprocal,hu2023broadband}. The generalization to one- and two-dimensional resonator networks can enhance the density of integration and provide new possible platforms for the realization of topological photonics with band structures and protect edge or entangled states~\cite{dai2022topologically,hashemi2025topological}.

\section{Conclusion} \label{Sec:conclusion}
In summary, we have delved into the photon transport and blockade, particularly focusing on the non-Markovian interactions between a microring resonator and a waveguide. By employing the two-point coupling approach from the perspective of scattering matrices, we have uncovered how a time delay can be induced by the propagating photons in the waveguide, subsequently impacting both single-photon and two-photon transports.
As a result, by tuning the waveguide length between two coupled points, the two input photons cannot be ideally transmitted, leading to the manifestation of photon blockade. 
Furthermore, we have also explored the generation of photon states within the resonator by applying a classical drive upon the resonator. Through this approach, we have discerned that the occurrence of photon blockade, as well as the modulation of photon number states, can be achieved by tuning the delayed phase in the non-Markovian system. Our research explores the intricate dynamics of photon transport and blockade in the framework of non-Markovian interactions between a microring resonator and a waveguide. This exploration presents intriguing possibilities for further exploration and application in the realm of quantum optics, quantum networking, and room-temperature quantum information processing.

\emph{Acknowledgement.---}The author thanks Chang-Ling Zou, Ming Li, Cong-Hua Yan and Re-Bing Wu for helpful discussions and suggestions. 

\begin{widetext}
\appendix

\section{Derivation on non-Markovian single-photon transport} \label{Sec:AppendixSinglephoton}
Here we derive the scattering matrix for the single-photon transport based on its definition in Eq.~(\ref{con:SDefineSinglePhoton}) in the main text.

When $g_1 \neq 0$ and  $g_2 \neq 0$, the scattering matrix for single-photon transport satisfies that
\begin{equation} \label{con:OnePTsite2}
\begin{aligned}
\left\langle \nu \right| S_1 \left|\omega \right\rangle = &\left\langle 0 \right|  \frac{1}{\sqrt{2\pi}} \int_{-\infty}^t b_{\rm out}(t) e^{i\nu t}\mathrm{d}t B_{\rm in}^{\dag}(\omega) \left| 0 \right\rangle\\
=& \left\langle 0 \right| B_{\rm in}(\nu) B_{\rm in}^{\dag}(\omega) \left| 0 \right\rangle + \sqrt{\frac{g_1}{2\pi}} \left\langle 0 \right| \int_{-\infty}^t  a\left( t - \frac{x}{c}\right) e^{i\nu t}\mathrm{d}t B_{\rm in}^{\dag}(\omega) \left| 0 \right\rangle \\
&+\sqrt{\frac{g_2}{2\pi}}  \left\langle 0 \right| \int_{-\infty}^t a\left( t - \frac{x-L}{c}\right) e^{i\nu t}\mathrm{d}t B_{\rm in}^{\dag}(\omega) \left| 0 \right\rangle.
\end{aligned}
\end{equation}

We denote 
\begin{equation} \label{con:GdefineTwopoint}
\begin{aligned}
\mathcal{A}(t) = \sqrt{\frac{g_1}{2\pi}}\left\langle 0 \right| a\left( t - \frac{x}{c}\right) \left| \omega^+ \right\rangle + \sqrt{\frac{g_2}{2\pi}}  \left\langle 0 \right| a\left( t - \frac{x-L}{c}\right) \left| \omega^+ \right\rangle,
\end{aligned}
\end{equation}
then according to Eq.~(\ref{con:adynamics}) in the main text,
\begin{equation} \label{con:OnePTsite3}
\begin{aligned}
\dot{\mathcal{A}}(t)=&\left\langle 0 \right| \left[-\sqrt{\frac{g_1}{2\pi}} \left(i \omega_a   + \kappa_a\right)  a\left( t - \frac{x}{c}\right) - \sqrt{\frac{g_1g_2}{2\pi}} b_{\rm in}\left(t-  \frac{L}{c}\right) - \frac{g_1}{\sqrt{2\pi}} b_{\rm in} \left(t\right)  \right]\left| \omega^+ \right\rangle\\
& - \left\langle 0 \right| \left[ \sqrt{\frac{g_2}{2\pi}} \left(i \omega_a   + \kappa_a\right)   a\left( t - \frac{x-L}{c}\right) + \sqrt{\frac{g_1g_2}{2\pi}}b_{\rm in}\left(t+  \frac{L}{c}\right) + \frac{g_2}{\sqrt{2\pi}} b_{\rm in}(t) \right]\left| \omega^+ \right\rangle,
\end{aligned}
\end{equation} 
where
\begin{equation} \label{con:bintminomega}
\begin{aligned}
&~~~~ \left\langle 0 \right| b_{\rm in}\left(t- \frac{L}{c}\right)  \left| \omega^+ \right\rangle\\
&= \left[ b_{\rm in}^{\dag}\left(t- \frac{L}{c}\right)  \left| 0 \right \rangle \right]^{\rm T} \left| \omega^+ \right\rangle\\
&= \left[ \frac{1}{\sqrt{2\pi}} \int_{-\infty}^{\infty}  B_{\rm in}^{\dag}(\omega') e^{i\omega' \left(t- \frac{L}{c}\right)}\mathrm{d}\omega'  \left| 0 \right \rangle \right]^{\rm T} \left| \omega^+ \right\rangle\\
&=  \frac{1}{\sqrt{2\pi}} \int_{-\infty}^{\infty}  e^{-i\omega' \left(t- \frac{L}{c}\right)} \left\langle \omega' \right |  \left| \omega^+ \right\rangle \mathrm{d}\omega' \\
&=  \frac{1}{\sqrt{2\pi}} e^{-i\omega \left(t- \frac{L}{c}\right)},
\end{aligned}
\end{equation}
and similarly~\cite{fan2010input}
\begin{equation} \label{con:bintPlusomega}
\begin{aligned}
& \left\langle 0 \right| b_{\rm in}\left(t+ \frac{L}{c}\right)  \left| \omega^+ \right\rangle=  \frac{1}{\sqrt{2\pi}}  e^{-i\omega \left(t+ \frac{L}{c}\right)}.
\end{aligned}
\end{equation} 

Based on  the following Fourier transforms for $\mathcal{A}(t)$ and $\dot{\mathcal{A}}(t)$,
\begin{subequations}
\begin{numcases}{}
\mathcal{F}_{\mathcal{A}}(\nu) =\frac{1}{\sqrt{2\pi}}\int_{-\infty}^{\infty} \mathcal{A}(t) e^{i\nu t}\mathrm{d}t,\\
(-i\nu)\mathcal{F}_{\mathcal{A}}(\nu) = \frac{1}{\sqrt{2\pi}}\int_{-\infty}^{\infty} \dot{\mathcal{A}}(t) e^{i\nu t}\mathrm{d}t,
\end{numcases}
\end{subequations}
and in combination with Eq.~(\ref{con:OnePTsite3}), we have
\begin{equation} \label{con:FTrelationTwoPointSolve}
\begin{aligned}
&\mathcal{F}_{\mathcal{A}}(\nu) = - \frac{ g_1 + g_2 + 2\sqrt{g_1g_2}  \cos(\omega L/c)}{\kappa_a +i\left(\omega_a -\omega\right) }\delta(\omega- \nu) .
\end{aligned}
\end{equation} 

Thus the transport of single photon can be evaluated by the scattering matrix as
\begin{equation} \label{con:SSinglePhotonResultFinalAppendix}
\begin{aligned}
\left\langle \nu \right| S_1 \left|\omega \right\rangle = \delta\left(\omega-\nu \right) 
- \frac{ g_1 + g_2 + 2\sqrt{g_1g_2}  \cos(\omega L/c)}{\kappa_a +i\left(\omega_a -\omega\right) }\delta(\omega- \nu).
\end{aligned}
\end{equation}

\section{Derivation on non-Markovian two-photon transport} \label{Sec:AppendixTwophoton}
According to the definition of scattering matrix in Eq.~(\ref{con:TwoPhotonMatrix}) in the main text, the two input photons and the two output photons are not distinguished. We can solve the scattering matrix for two-photon transport by inserting an identity operator as
\begin{equation} \label{con:TwoPhotonMatrixAppendix}
\begin{aligned}
&\left\langle \nu_1 \nu_2 \right| S_2 \left|\omega_1\omega_2 \right\rangle = \left\langle 0 \right| B_{\rm out}\left(\nu_1\right) \int_{-\infty}^{\infty}  \left| \omega^+ \right\rangle  \left\langle \omega^+ \right| \mathrm{d}\omega B_{\rm out}\left(\nu_2\right) B_{\rm in}^{\dag}\left(\omega_1\right) B_{\rm in}^{\dag}\left(\omega_2\right) \left| 0 \right\rangle,
\end{aligned}
\end{equation}  
where the former component can be solved as
\begin{equation} \label{con:InsertCal1}
\begin{aligned}
 &~~~~\left\langle 0 \right| B_{\rm out}\left(\nu_1\right) \int_{-\infty}^{\infty}   \left| \omega^+ \right\rangle  \left\langle \omega^+ \right| \mathrm{d}\omega \\
&=\int_{-\infty}^{\infty}  \left\langle 0 \right| B_{\rm out}\left(\nu_1\right) B_{\rm in}^{\dag}(\omega) |0\rangle  \left\langle \omega^+ \right| \mathrm{d}\omega \\ 
& =\int_{-\infty}^{\infty} t_{\omega}\delta\left(\omega-\nu_1 \right) \left\langle \omega^+ \right| \mathrm{d}\omega,
\end{aligned}
\end{equation} 
which is based on the scattering matrix for single-photon transport in Eq.~(\ref{con:SSinglePhotonResultFinal}) or Appendix~\ref{Sec:AppendixSinglephoton}. Namely a single-photon transport process is included in the two-photon transport as
\begin{equation}
\begin{aligned}
\left\langle 0 \right| B_{\rm out}\left(\nu_1\right)\int_{-\infty}^{\infty}  \left| \omega^+ \right\rangle  \left\langle \omega^+ \right| \mathrm{d}\omega =  t_{\nu_1} \left\langle \nu_1^+ \right|,
\end{aligned}
\end{equation} 
and Eq.~(\ref{con:TwoPhotonMatrixAppendix}) can be further simplified as
\begin{equation} \label{con:TwoPhotonMatrixCal2}
\begin{aligned}
\left\langle \nu_1 \nu_2 \right|S_2 \left|\omega_1\omega_2 \right\rangle = t_{\nu_1} \left \langle \nu_1^+ \right| B_{\rm out}\left(\nu_2\right) B_{\rm in}^{\dag}\left(\omega_1\right) B_{\rm in}^{\dag}\left(\omega_2\right) \left| 0 \right\rangle.
\end{aligned}
\end{equation}  

In combination with Eq.~(\ref{boutIO}) and the Fourier inversions of Eqs.~(\ref{con:bintAinOmegaRelationCal2},\ref{con:bouttAoutOmegaRelation})
\begin{subequations} \label{eq:AoutRelation}
\begin{numcases}{}
B_{\rm in}(\omega) = \frac{1}{\sqrt{2\pi}} \int_{t_0}^{\infty}  b_{\rm in}(t)  e^{i\omega t} \mathrm{d}t ,\\
B_{\rm out}(\omega) = \frac{1}{\sqrt{2\pi}} \int_{t_0}^{\infty} b_{\rm out}(t)  e^{i\omega t}\mathrm{d}t,
\end{numcases}
\end{subequations}
we have
\begin{equation} \label{con:AoutW2}
\begin{aligned} 
B_{\rm out}\left(\nu_2\right) &= \frac{1}{\sqrt{2\pi}} \int_{t_0}^{\infty} b_{\rm out}(t)  e^{i\nu_2 t}\mathrm{d}t \\
&=B_{\rm in}\left(\nu_2\right) + \int_{t_0}^{\infty}  \left[ \sqrt{\frac{g_1}{2\pi}} a\left( t - \frac{x}{c}\right) +  \sqrt{\frac{g_2}{2\pi}}  a\left(t - \frac{x-L}{c} \right) \right] e^{i\nu_2 t} \mathrm{d}t.
\end{aligned}
\end{equation} 

Now we take the second line of Eq.~(\ref{con:AoutW2}) into Eq.~(\ref{con:TwoPhotonMatrixCal2}). On one hand, using the commutative relationship in Eq.~(\ref{eq:AinOmegaCommut}) in the main text, we can rewrite the component at the righthand side (RHS) of Eq.~(\ref{con:TwoPhotonMatrixCal2}) determined by $B_{\rm in}\left(\nu_2\right)$ as~\cite{fan2010input}
\begin{equation} \label{con:TwoPhotonMatrixCal3Part1}
\begin{aligned}
t_{\nu_1} \left \langle \nu_1^+ \right| B_{\rm in}\left(\nu_2\right) B_{\rm in}^{\dag}\left(\omega_1\right) B_{\rm in}^{\dag}\left(\omega_2\right) \left| 0 \right\rangle&=t_{\nu_1} \left \langle \nu_1^+ \right| B_{\rm in}^{\dag}\left(\omega_1\right) B_{\rm in}\left(\nu_2\right) \left| \omega_2^+ \right\rangle + t_{\nu_1}  \delta\left( \nu_2 - \omega_1\right) \left \langle \nu_1^+ \right|  \left| \omega_2^+ \right\rangle\\
&=t_{\nu_1} \delta\left( \nu_1 - \omega_1\right) \delta \left( \nu_2 - \omega_2\right)  + t_{\nu_1}  \delta\left( \nu_2 - \omega_1\right) \delta\left( \nu_1 - \omega_2\right).
\end{aligned}
\end{equation}  

On the other hand, for the remaining components at the RHS of Eq.~(\ref{con:TwoPhotonMatrixCal2}) with two input photons with frequencies $\omega_1$ and $\omega_2$, we denote
\begin{subequations} \label{eq:K1K2}
\begin{numcases}{}
\mathcal{K}\left(t,\nu_1\right) =  \left \langle \nu_1^+ \right| a\left( t - \frac{x}{c}\right) B_{\rm in}^{\dag}\left(\omega_1\right) B_{\rm in}^{\dag}\left(\omega_2\right) \left| 0 \right\rangle , \label{TwophoK1}\\
\mathcal{K}\left(t+\frac{L}{c},\nu_1\right) =  \left \langle \nu_1^+ \right| a\left( t - \frac{x-L}{c}\right) B_{\rm in}^{\dag}\left(\omega_1\right) B_{\rm in}^{\dag}\left(\omega_2\right) \left| 0 \right\rangle,\label{TwophoK2}\\
\mathcal{F}\left(\nu_1,\nu_2\right) = \frac{1}{\sqrt{2\pi}} \int_{t_0}^{\infty}  \mathcal{K}\left(t,\nu_1\right) e^{i\nu_2 t} \mathrm{d}t,\label{TwophoF1}
\end{numcases}
\end{subequations}
then Eq.~(\ref{con:TwoPhotonMatrixCal2}) can be rewritten as
\begin{equation} \label{con:TwoPhotonMatrixCal3V2}
\begin{aligned}
\left\langle \nu_1 \nu_2 \right| S_2 \left|\omega_1\omega_2 \right\rangle= &t_{\nu_1} \left \langle \nu_1^+ \right| B_{\rm in}\left(\nu_2\right) B_{\rm in}^{\dag}\left(\omega_1\right) B_{\rm in}^{\dag}\left(\omega_2\right) \left| 0 \right\rangle +  t_{\nu_1} \left(\sqrt{g_1}+ \sqrt{g_2}e^{-i\nu_2 L/c} \right)\mathcal{F}\left(\nu_1,\nu_2\right).
\end{aligned}
\end{equation}

Combined with the operator dynamics in Eq.~(\ref{con:adynamics}), the derivative of $\mathcal{K}\left(t,\nu_1\right)$ upon time $t$ can be derived as
\begin{equation} \label{con:dotK1}
\begin{aligned}
\dot{\mathcal{K}}\left(t,\nu_1\right)   =& -\left(i \omega_a+ \kappa_a \right) \mathcal{K}\left(t,\nu_1\right)  -2i\chi \left \langle \nu_1^+ \right|  a^{\dag}\left( t - \frac{x}{c}\right)a^2\left( t - \frac{x}{c}\right) B_{\rm in}^{\dag}\left(\omega_1\right) B_{\rm in}^{\dag}\left(\omega_2\right) \left| 0 \right\rangle \\
&-  \left \langle \nu_1^+ \right|  \left [\sqrt{g_1} b_{\rm in} \left(t\right) + \sqrt{g_2} b_{\rm in}\left(t-  \frac{L}{c}\right)\right ] B_{\rm in}^{\dag}\left(\omega_1\right) B_{\rm in}^{\dag}\left(\omega_2\right) \left| 0 \right\rangle.
\end{aligned}
\end{equation}  

Based on this, we derive the scattering matrix for two-photon circumstance as follows by separating the calculations into three parts.

(\romannumeral 1) For the second component at the RHS of Eq.~(\ref{con:dotK1}) influenced by the interactions with the Kerr material, we notice that~\cite{SmatrixCal}
\begin{equation} \label{con:dotK1Kerr}
\begin{aligned}
&~~~~ \chi\left \langle \nu_1^+ \right|  a^{\dag}\left( t - \frac{x}{c}\right)a^2\left( t - \frac{x}{c}\right) B_{\rm in}^{\dag}\left(\omega_1\right) B_{\rm in}^{\dag}\left(\omega_2\right) \left| 0 \right\rangle \\
&=\chi\left \langle \nu_1^+ \right|  a^{\dag}\left( t - \frac{x}{c}\right)a\left( t - \frac{x}{c}\right)  \int_{-\infty}^{\infty}  \left| \omega^+ \right\rangle  \left\langle \omega^+ \right| \mathrm{d}\omega  a\left( t - \frac{x}{c}\right) B_{\rm in}^{\dag}\left(\omega_1\right) B_{\rm in}^{\dag}\left(\omega_2\right) \left| 0 \right\rangle \\
&=\chi \int_{-\infty}^{\infty} \left \langle \nu_1^+ \right|  a^{\dag}\left( t - \frac{x}{c}\right) |0\rangle \langle 0| a\left( t - \frac{x}{c}\right)    \left| \omega^+ \right\rangle  \left\langle \omega^+ \right| a\left( t - \frac{x}{c}\right)B_{\rm in}^{\dag}\left(\omega_1\right) B_{\rm in}^{\dag}\left(\omega_2\right) \left| 0 \right\rangle \mathrm{d}\omega,
\end{aligned}
\end{equation}  
where we insert the basis $|0\rangle \langle 0|$ due to the annihilation of two photons by the operator $a^2\left( t - x/c\right)$.

According to Eq.~(\ref{con:GdefineTwopoint}) for single-photon transport in Appendix~\ref{Sec:AppendixSinglephoton}, we denote $\mathcal{A}_1(t,\omega) =  \left\langle 0 \right| a\left( t - x/c\right) \left| \omega^+ \right\rangle$, $\mathcal{A}_1^*(t,\omega) =  \left \langle \omega^+ \right|  a^{\dag}\left( t - x/c\right) |0\rangle $, and
\begin{equation} \label{con:dotK1KerrPart1}
\begin{aligned}
\dot{\mathcal{A}}_1(t,\omega) &= \langle 0| \dot{a}\left( t - \frac{x}{c}\right)    \left| \omega^+ \right\rangle \\
& = -\left( \kappa_a +  i \omega_a\right) \mathcal{A}_1(t,\omega)-  \left( \sqrt{g_1}+ \sqrt{g_2} e^{i\omega \frac{L}{c}} \right)e^{-i\omega t},
\end{aligned}
\end{equation}
which is independent from the two-photon process~\cite{SmatrixCal}. Assisted by the dynamics with a time-varying envelope $\mathcal{A}_1(t,\omega) e^{\left(\kappa_a +i\omega_a\right) t} $, 
$\mathcal{A}_1(t,\omega)$ can be solved as
\begin{equation} \label{con:dotK1KerrPart1SimplifySolution}
\begin{aligned}
\mathcal{A}_1(t,\omega) & = -\frac{\sqrt{g_1} + \sqrt{g_2} e^{i\omega \frac{L}{c}}}{\kappa_a+i \left(\omega_a-\omega\right)  }e^{-i\omega t} \triangleq -\Gamma(\omega)e^{-i\omega t},
\end{aligned}
\end{equation}
with $\Gamma(\omega)$ defined in Eq.~(\ref{con:GammaDefine}) in the main text.

Take the solution of $\mathcal{A}_1(t,\omega)$ into Eq.~(\ref{con:dotK1Kerr}), we have
\begin{equation} \label{con:dotK1KerrCaltomegaSimplify}
\begin{aligned}
&~~~~\chi \left \langle \nu_1^+ \right|  a^{\dag}\left( t - \frac{x}{c}\right)a^2\left( t - \frac{x}{c}\right) B_{\rm in}^{\dag}\left(\omega_1\right) B_{\rm in}^{\dag}\left(\omega_2\right) \left| 0 \right\rangle \\
&= \chi \int_{-\infty}^{\infty} \mathcal{A}_1^*\left(t,\nu_1\right) \mathcal{A}_1(t,\omega) \left\langle \omega^+ \right| a\left( t - \frac{x}{c}\right)B_{\rm in}^{\dag}\left(\omega_1\right) B_{\rm in}^{\dag}\left(\omega_2\right) \left| 0 \right\rangle \mathrm{d}\omega\\
& = \chi \int_{-\infty}^{\infty} \frac{\sqrt{g_1} + \sqrt{g_2} e^{-i\nu_1 \frac{L}{c}}}{\kappa_a+i \left(\nu_1-\omega_a\right)  }\frac{\sqrt{g_1} + \sqrt{g_2} e^{i\omega \frac{L}{c}}}{\kappa_a-i \left(\omega-\omega_a\right)  }e^{i\left(\nu_1-\omega\right) t}  \left\langle \omega^+ \right| a\left( t - \frac{x}{c}\right)B_{\rm in}^{\dag}\left(\omega_1\right) B_{\rm in}^{\dag}\left(\omega_2\right) \left| 0 \right\rangle \mathrm{d}\omega\\
& =  \chi  \Gamma^*\left(\nu_1\right) \int_{-\infty}^{\infty}\Gamma\left(\omega\right) e^{i\left(\nu_1-\omega\right) t}  \left\langle \omega^+ \right| a\left( t - \frac{x}{c}\right)B_{\rm in}^{\dag}\left(\omega_1\right) B_{\rm in}^{\dag}\left(\omega_2\right) \left| 0 \right\rangle \mathrm{d}\omega.
\end{aligned}
\end{equation}

To derive $\mathcal{F}\left(\nu_1,\nu_2\right)$ according to its definition in Eq.~(\ref{TwophoF1}), we do the Fourier transform with the frequency $\nu_2$ upon the two sides of Eq.~(\ref{con:dotK1}), and that for the second component at the RHS can be represented as a function of $\nu_1$ and $\nu_2$ according to Eq.~(\ref{con:dotK1KerrCaltomegaSimplify}) as
\begin{equation} 
\begin{aligned}\label{con:defT}
 \mathcal{T}\left(\nu_1,\nu_2 \right) =&- \frac{2i\chi}{\sqrt{2\pi}}\int_{t_0}^{\infty}\left \langle \nu_1^+ \right|  a^{\dag}\left( t - \frac{x}{c}\right)a^2\left( t - \frac{x}{c}\right) B_{\rm in}^{\dag}\left(\omega_1\right) B_{\rm in}^{\dag}\left(\omega_2\right) \left| 0 \right\rangle e^{i\nu_2 t} \mathrm{d}t\\
=&- \frac{2i\chi}{\sqrt{2\pi}}\Gamma^*\left(\nu_1\right)\int_{-\infty}^{\infty} \int_{t_0}^{\infty}   \Gamma\left(\omega\right) e^{i\left(\nu_1+\nu_2-\omega\right) t}  \left\langle \omega^+ \right| a\left( t - \frac{x}{c}\right)B_{\rm in}^{\dag}\left(\omega_1\right) B_{\rm in}^{\dag}\left(\omega_2\right) \left| 0 \right\rangle  \mathrm{d}t\mathrm{d}\omega\\
=&- \frac{2i\chi}{\sqrt{2\pi}}\Gamma^*\left(\nu_1\right)   \int_{-\infty}^{\infty} \Gamma\left(\omega\right) \int_{t_0}^{\infty}\mathcal{K}\left(t,\omega\right)  e^{i\left(\nu_1+\nu_2-\omega\right) t}  \mathrm{d}t\mathrm{d}\omega.
\end{aligned}
\end{equation}  

Generalized by Eq.~(\ref{TwophoF1}), 
\begin{equation} 
\begin{aligned}
\frac{1}{\sqrt{2\pi}} \int_{t_0}^{\infty}   \mathcal{K}\left(t,\omega\right)e^{i\left(\nu_1+\nu_2-\omega\right) t} \mathrm{d}t = \mathcal{F}\left(\omega,\nu_1+\nu_2-\omega\right),
\end{aligned}
\end{equation}  
then in Eq.~(\ref{con:defT}), 
\begin{equation} 
\begin{aligned}\label{con:defTV2}
\mathcal{T}\left(\nu_1,\nu_2 \right)& = -2i\chi  \Gamma^*\left(\nu_1\right) \int_{-\infty}^{\infty}\Gamma\left(\omega\right)  \mathcal{F}\left(\omega,\nu_1+\nu_2-\omega\right)\mathrm{d}\omega.
\end{aligned}
\end{equation}

(\romannumeral 2) For the third component at the RHS of Eq.~(\ref{con:dotK1}), considering that
\begin{equation} 
\begin{aligned}
B_{\rm in}(\omega) & = \frac{1}{\sqrt{2\pi}} \int_{-\infty}^{\infty} b_{\rm in}(t) e^{i\omega t}\mathrm{d}t,
\end{aligned}
\end{equation}
as an inverse format of Eq.~(\ref{con:bintAinOmegaRelationCal2}) in the main text, then 
\begin{equation} 
\begin{aligned}
&\frac{1}{\sqrt{2\pi}} \int_{-\infty}^{\infty} \left [\sqrt{g_1} b_{\rm in} \left(t\right) + \sqrt{g_2} b_{\rm in}\left(t-  \frac{L}{c}\right)\right ] e^{i\nu_2 t}\mathrm{d}\omega=\left(\sqrt{g_1}+ \sqrt{g_2}e^{i \nu_2\frac{L}{c}} \right)B_{\rm in}\left(\nu_2\right). 
\end{aligned}
\end{equation}

According to Eq.~(\ref{con:dotK1}), given $\nu_1$ as the frequency of the first output photon, we have
\begin{equation} \label{con:dotK1FreqDomain2}
\begin{aligned}
\left [ \kappa_a + i\left(\omega_a - \nu_2\right) \right]\mathcal{F}\left(\nu_1,\nu_2\right)
=  \mathcal{T}\left(\nu_1,\nu_2 \right) -\left(\sqrt{g_1}+ \sqrt{g_2}e^{i \nu_2\frac{L}{c}} \right)  \left \langle \nu_1^+ \right| B_{\rm in}\left(\nu_2\right) B_{\rm in}^{\dag}\left(\omega_1\right) B_{\rm in}^{\dag}\left(\omega_2\right) \left| 0 \right\rangle, 
\end{aligned}
\end{equation}  
then 
\begin{equation} \label{con:FK1SolutionResult}
\begin{aligned}
\mathcal{F}\left(\nu_1,\nu_2\right)  =& \frac{\mathcal{T}\left(\nu_1,\nu_2 \right)}{\kappa_a + i\left(\omega_a - \nu_2\right) } - \frac{\sqrt{g_1} + \sqrt{g_2} e^{i\nu_2\frac{L}{c}}}{ \kappa_a + i\left(\omega_a - \nu_2\right) } \left[\delta\left( \nu_1 - \omega_1\right) \delta \left( \nu_2 - \omega_2\right)  +  \delta\left( \nu_2 - \omega_1\right) \delta\left( \nu_1 - \omega_2\right)\right], 
\end{aligned}
\end{equation}  
with $\mathcal{T}\left(\nu_1,\nu_2 \right)$ given by Eq.~(\ref{con:defTV2}).

Then for the RHS of Eq.~(\ref{con:TwoPhotonMatrixCal3V2}), we have
\begin{equation} \label{con:TwoPhotonMatrixCal3V3}
\begin{aligned}
&t_{\nu_1} \left \langle \nu_1^+ \right| B_{\rm in}\left(\nu_2\right) B_{\rm in}^{\dag}\left(\omega_1\right) B_{\rm in}^{\dag}\left(\omega_2\right) \left| 0 \right\rangle +  t_{\nu_1} \left(\sqrt{g_1}+ \sqrt{g_2}e^{-i\nu_2 L/c} \right)\mathcal{F}\left(\nu_1,\nu_2\right)\\
=&t_{\nu_1} \left[\delta\left( \nu_1 - \omega_1\right) \delta \left( \nu_2 - \omega_2\right)  +   \delta\left( \nu_2 - \omega_1\right) \delta\left( \nu_1 - \omega_2\right) \right] + t_{\nu_1}  \frac{\sqrt{g_1}+ \sqrt{g_2}e^{-i\nu_2 L/c}}{\kappa_a + i\left(\omega_a - \nu_2\right) }\mathcal{T}\left(\nu_1,\nu_2 \right) \\
&- t_{\nu_1}  \frac{g_1 + g_2 +2 \sqrt{g_1 g_2} \cos\left(\nu_2 L/c\right)}{ \kappa_a + i\left(\omega_a - \nu_2\right) } \left[\delta\left( \nu_1 - \omega_1\right) \delta \left( \nu_2 - \omega_2\right)  +  \delta\left( \nu_2 - \omega_1\right) \delta\left( \nu_1 - \omega_2\right)\right] \\
=&t_{\nu_1}t_{\nu_2} \left[ \delta\left( \nu_1 - \omega_1\right) \delta \left( \nu_2 - \omega_2\right)  +   \delta\left( \nu_2 - \omega_1\right) \delta\left( \nu_1 - \omega_2\right) \right]+ t_{\nu_1}  \frac{\sqrt{g_1}+ \sqrt{g_2}e^{-i\nu_2 L/c}}{\kappa_a + i\left(\omega_a - \nu_2\right) }\mathcal{T}\left(\nu_1,\nu_2 \right).
\end{aligned}
\end{equation}

(\romannumeral 3) Based on Eq.~(\ref{con:defTV2}) and Eq.~(\ref{con:FK1SolutionResult}), we solve $\mathcal{T}\left(\nu_1,\nu_2\right)$ as follows~\cite{SmatrixCal}. 

By replacing $\nu_2$ with $\nu_2-\nu_1$ in Eq.~(\ref{con:FK1SolutionResult}), we have
\begin{equation}  \label{con:GdefineV2}
\begin{aligned}
\mathcal{F}\left(\nu_1,\nu_2-\nu_1\right) =&-2i\chi  \frac{ \Gamma^*\left(\nu_1\right)}{\kappa_a + i\left(\omega_a - \nu_2 + \nu_1\right) } \mathcal{G}\left( \nu_2 \right)  \\
&- \frac{\sqrt{g_1} + \sqrt{g_2} e^{i\frac{\left(\nu_2-\nu_1 \right)L}{c}}}{ \kappa_a + i\left(\omega_a - \nu_2+\nu_1\right) } \left[\delta\left( \nu_1 - \omega_1\right) \delta \left( \nu_2-\nu_1- \omega_2\right)  +  \delta\left(\nu_2-\nu_1 - \omega_1\right) \delta\left( \nu_1 - \omega_2\right)\right],
\end{aligned}
\end{equation}
and generalized from Eq.~(\ref{con:defTV2}),
\begin{equation} 
\begin{aligned}
\mathcal{T}\left(\nu_1,\nu_2-\nu_1 \right) = -2i\chi  \Gamma^*\left(\nu_1\right)\int_{-\infty}^{\infty}\Gamma\left(\omega\right)  \mathcal{F}\left(\omega,\nu_2-\omega\right)\mathrm{d}\omega = -2i\chi  \Gamma^*\left(\nu_1\right)\mathcal{G}\left( \nu_2 \right),
\end{aligned}
\end{equation}
with
\begin{equation} \label{con:Gdefine}
\begin{aligned}
\mathcal{G}\left( \nu_2 \right) =  \int_{-\infty}^{\infty}\Gamma\left(\omega\right)  \mathcal{F}\left(\omega,\nu_2-\omega\right)\mathrm{d}\omega.
\end{aligned}
\end{equation}

By replacing $\nu_1$ with $\omega$ in Eq.~(\ref{con:GdefineV2}),
\begin{equation}  \label{con:FcalGnv2}
\begin{aligned}
\mathcal{F}\left(\omega,\nu_2-\omega\right)=&-2i\chi  \frac{ \Gamma^*\left(\omega\right)}{\kappa_a + i\left(\omega_a - \nu_2 + \omega\right) } \mathcal{G}\left( \nu_2 \right)  \\
&- \frac{\sqrt{g_1} + \sqrt{g_2} e^{i\frac{\left(\nu_2-\omega \right)L}{c}}}{ \kappa_a + i\left(\omega_a - \nu_2+\omega\right) } \left[\delta\left( \omega - \omega_1\right) \delta \left( \nu_2-\omega- \omega_2\right)  +  \delta\left(\nu_2-\omega - \omega_1\right) \delta\left( \omega - \omega_2\right)\right],
\end{aligned}
\end{equation}
and Eq.~(\ref{con:Gdefine}) can be further derived as
\begin{equation} \label{con:Gcal1}
\begin{aligned}
\mathcal{G}\left( \nu_2 \right) = &-2i\chi  \mathcal{G}\left( \nu_2 \right) \int_{-\infty}^{\infty}  \frac{ \Gamma\left(\omega\right) \Gamma^*\left(\omega\right)}{\kappa_a + i\left(\omega_a - \nu_2 + \omega\right) } \mathrm{d}\omega\\
&- \int_{-\infty}^{\infty}\Gamma\left(\omega\right) \frac{\sqrt{g_1} + \sqrt{g_2} e^{i\frac{\left(\nu_2-\omega \right)L}{c}}}{ \kappa_a + i\left(\omega_a - \nu_2+\omega\right) } \left[\delta\left( \omega - \omega_1\right) \delta \left( \nu_2-\omega- \omega_2\right)  +  \delta\left(\nu_2-\omega - \omega_1\right) \delta\left( \omega - \omega_2\right)\right]  \mathrm{d}\omega\\
=&-2i\chi  \mathcal{G}\left( \nu_2 \right) \int_{-\infty}^{\infty}  \frac{ \Gamma\left(\omega\right) \Gamma^*\left(\omega\right)}{\kappa_a + i\left(\omega_a - \nu_2 + \omega\right) } \mathrm{d}\omega - \left[\Gamma\left(\omega_1\right) \Gamma\left(\nu_2-\omega_1\right) +  \Gamma\left(\omega_2\right)  \Gamma\left(\nu_2-\omega_2\right)\right] \delta\left(\nu_2-\omega_1 - \omega_2\right) .
\end{aligned}
\end{equation}

According to the method relied on the residue theorem adopted in Ref.~\cite{SmatrixCal},
\begin{equation} \label{con:residue}
\begin{aligned}
\int_{-\infty}^{\infty}  \frac{ \Gamma\left(\omega\right) \Gamma^*\left(\omega\right)}{\kappa_a + i\left(\omega_a - \nu_2 + \omega\right) } \mathrm{d}\omega & \approx \frac{\pi \left(g_1 + g_2\right)  }{\kappa_a \left[2\kappa_a +i \left(2\omega_a - \nu_2 \right) \right]},
\end{aligned}
\end{equation}
by neglecting the oscillating zero-mean terms.

Then Eq.~(\ref{con:Gcal1}) can be rewritten as
\begin{equation} \label{con:Gcal2}
\begin{aligned}
\mathcal{G}\left( \nu_2 \right) =&-2i\chi  \mathcal{G}\left( \nu_2 \right) \frac{\pi \left(g_1 + g_2\right) }{\kappa_a \left[2\kappa_a +i \left(2\omega_a - \nu_2 \right) \right]}\\
& - \Gamma\left(\omega_1\right) \Gamma\left(\nu_2-\omega_1\right) \delta \left( \nu_2-\omega_1- \omega_2\right) -  \Gamma\left(\omega_2\right)  \Gamma\left(\nu_2-\omega_2\right) \delta\left(\nu_2-\omega_2 - \omega_1\right) ,
\end{aligned}
\end{equation}
thus $\mathcal{G}\left( \nu_2 \right) $ can be solved as
\begin{equation} \label{con:Gcal3}
\begin{aligned}
\mathcal{G}\left( \nu_2 \right) 
=&-\frac{2\kappa_a \left[2\kappa_a +i \left(2\omega_a - \nu_2 \right) \right]\Gamma\left(\omega_1\right) \Gamma\left(\omega_2\right) }{\kappa_a \left[2\kappa_a +i \left(2\omega_a - \nu_2 \right) \right]+2\pi i\chi   \left(g_1 + g_2 \right)  }\delta \left( \nu_2-\omega_1- \omega_2\right).
\end{aligned}
\end{equation}

Namely in Eq.~(\ref{con:FcalGnv2}),
\begin{equation} \label{con:Fnvomega}
\begin{aligned}
\mathcal{F}\left(\omega,\nu_2-\omega\right)=&\frac{ 2i\chi \Gamma^*\left(\omega\right)}{\kappa_a + i\left(\omega_a - \nu_2 + \omega\right) }\frac{2\kappa_a \left[2\kappa_a +i \left(2\omega_a - \nu_2 \right) \right]\Gamma\left(\omega_1\right) \Gamma\left(\omega_2\right) \delta \left( \nu_2-\omega_1- \omega_2\right) }{\kappa_a \left[2\kappa_a +i \left(2\omega_a - \nu_2 \right) \right]+2\pi i\chi   \left(g_1 + g_2 \right) }\\
&- \frac{\sqrt{g_1} + \sqrt{g_2} e^{i\frac{\left(\nu_2-\omega \right)L}{c}}}{ \kappa_a + i\left(\omega_a - \nu_2+\omega\right) } \left[\delta\left( \omega - \omega_1\right) \delta \left( \nu_2-\omega- \omega_2\right)  +  \delta\left(\nu_2-\omega - \omega_1\right) \delta\left( \omega - \omega_2\right)\right],
\end{aligned}
\end{equation}
then the first component at the RHS of Eq.~(\ref{con:GdefineV2}) can be derived by replacing $\omega$ with $\nu_1$ in Eq.~(\ref{con:Fnvomega}) as
\begin{equation} \label{con:FreplaceCal2}
\begin{aligned}
&\frac{\mathcal{T}\left(\nu_1,\nu_2-\nu_1 \right)}{\kappa_a + i\left(\omega_a - \nu_2 + \nu_1\right) }= \frac{ 2i\chi \Gamma^*\left(\nu_1\right)}{\kappa_a + i\left(\omega_a - \nu_2 + \nu_1\right) } \frac{2\kappa_a \left[2\kappa_a +i \left(2\omega_a - \nu_2 \right) \right]\Gamma\left(\omega_1\right) \Gamma\left(\omega_2\right) \delta \left( \nu_2-\omega_1- \omega_2\right) }{\kappa_a \left[2\kappa_a +i \left(2\omega_a - \nu_2 \right) \right]+2\pi i\chi   \left(g_1 + g_2 \right) },
\end{aligned}
\end{equation}
due to the first component at the RHS of Eq.~(\ref{con:FK1SolutionResult}).

By replacing $\nu_2 - \nu_1$ with $\nu_2$ in Eq.~(\ref{con:FreplaceCal2}), we can derive that
\begin{equation} \label{con:Tnu1nu2}
\begin{aligned}
\mathcal{T}\left(\nu_1,\nu_2 \right) &=  \Gamma^*\left(\nu_1\right)  \frac{4i\chi\kappa_a \left[2\kappa_a +i \left(2\omega_a - \nu_1- \nu_2 \right) \right]\Gamma\left(\omega_1\right) \Gamma\left(\omega_2\right)  }{\kappa_a \left[2\kappa_a +i \left(2\omega_a - \nu_1 - \nu_2 \right) \right]+2\pi i\chi   \left(g_1 + g_2 \right) }\delta \left( \nu_1+\nu_2-\omega_1- \omega_2\right) .
\end{aligned}
\end{equation}

Similar to Eq.~(\ref{con:SinglePhotontw}), $\eta_{\omega}$ has been defined in Eq.~(\ref{con:EtaDefine}) in the main text, and equivalently
\begin{equation} 
\begin{aligned}
\eta_{\omega}&=\frac{\kappa_a-i \left(\omega_a-\omega\right) }{\kappa_a+i \left(\omega_a-\omega\right) },
\end{aligned}
\end{equation}
then in the last component of Eq.~(\ref{con:TwoPhotonMatrixCal3V3}), we can do the following replacement for simplification as
\begin{equation}
\begin{aligned}
&t_{\nu_1}  \frac{\sqrt{g_1}+ \sqrt{g_2}e^{-i\nu_2 L/c}}{\kappa_a + i\left(\omega_a - \nu_2\right) }\Gamma^*\left(\nu_1\right) =t_{\nu_1} \eta_{\nu_2}\Gamma^*\left(\nu_1\right)\Gamma^*\left(\nu_2\right).
\end{aligned}
\end{equation}

Due to the mathematical structure of the two-photon Hilbert space~\cite{ShiTaoPRA}, the scattering matrix for two-photon transport can be derived as Eq.~(\ref{con:TwoPhotonMatrixResultGeneral}) in the main text.

The scattering matrix for two-photon transport has a similar format compared with those in various quantum systems such as in Refs.~\cite{ShiTaoPRA,TwoPhotonKerr,li2020photon,shi2011two}. Specially for a simplified case that $\chi = 0$, 
\begin{equation} \label{con:chizero}
\begin{aligned}
\left\langle \nu_1 \nu_2 \right| S_2 \left|\omega_1\omega_2 \right\rangle=&t_{\nu_1}t_{\nu_2} \left[ \delta\left( \nu_1 - \omega_1\right) \delta \left( \nu_2 - \omega_2\right)  +   \delta\left( \nu_2 - \omega_1\right) \delta\left( \nu_1 - \omega_2\right) \right],
\end{aligned}
\end{equation}  
and this agrees with Refs.~\cite{fan2010input,li2020photon}.

\end{widetext}

\bibliography{Blockade}
\end{document}